\documentclass[prb,twocolumn]{revtex4}

\usepackage{amssymb}
\usepackage{color}
\usepackage{amsthm}
\usepackage{graphicx}
\usepackage{dcolumn}
\usepackage{bm}
\usepackage{subfigure}
\usepackage{amssymb}
\usepackage{verbatim}
\usepackage{amscd}
\usepackage{amssymb}
\usepackage{amsmath, amsfonts}
\usepackage{setspace}
\usepackage[]{graphicx}        
\usepackage{amsthm}
\usepackage{enumerate}

\newcommand{\tb}[1]{\textbf{#1}}

\newcommand{\ba}[1]{\begin{align*} #1 \end{align*}}

\theoremstyle{plain}

\theoremstyle{plain}

\theoremstyle{plain}
 
\theoremstyle{plain}

\theoremstyle{remark}

\theoremstyle{conjecture}

\theoremstyle{observation}

\theoremstyle{definition}

\theoremstyle{corollary}

\theoremstyle{definition}

\theoremstyle{definition}

\theoremstyle{assumption}

\theoremstyle{definition}

\theoremstyle{problem}

\theoremstyle{fact}

\begin{document}

\title{Exotic topological order in fractal spin liquids}
\author{Beni Yoshida}
\affiliation{Institute for Quantum Information and Matter, California Institute of Technology, Pasadena, California 91125, USA}

\date{\today}
\begin{abstract}
We present a large class of three-dimensional spin models that possess topological order with stability against local perturbations, but are beyond description of topological quantum field theory. Conventional topological spin liquids, on a formal level, may be viewed as condensation of string-like extended objects with discrete gauge symmetries, being at fixed points with continuous scale symmetries. In contrast, ground states of fractal spin liquids are condensation of highly-fluctuating fractal objects with certain algebraic symmetries, corresponding to limit cycles under real-space renormalization group transformations which naturally arise from discrete scale symmetries of underlying fractal geometries. A particular class of three-dimensional models proposed in this paper may potentially saturate quantum information storage capacity for local spin systems.
\end{abstract}
\maketitle

\section{Introduction}

In a quantum many-body system at zero temperature, topological order may arise when a gapped ground state possesses long-range entanglement that cannot be detected by any local measurement or local order parameter.~\cite{Kitaev03} Its ground state properties are stable against any types of local perturbations, regardless of symmetries of perturbations, and depend only on global properties of geometric manifold on which the whole system is supported.~\cite{Wen90} The discovery of topological order, such as fractional quantum Hall systems~\cite{Tsui82, Laughlin83}, came as a great surprise as they are beyond description of the Landau-Ginzburg theory which was once believed to be \emph{the} ultimate theory of a quantum many-body system. It is now widely believed that the notion of topological order is essential in understanding the emergence of quantum phases with no local order in gapped quantum spin liquids, as seen in some frustrated anti-ferromagnets.~\cite{Wen89, Read91, Wen91, Anderson87, AKLT87, Senthil00, Moessner01, Kitaev06b} The study of topologically ordered systems is also of practical importance as they are physically natural platforms for realizations of fault-tolerant quantum information processing.~\cite{Kitaev03}

For topologically ordered spin liquids with discrete gauge symmetries, their low energy behavior is relatively well understood on a formal level as they are effectively described by topological quantum field theory (TQFT)~\cite{Wen90, Read91}, a field theory with invariance under continuous deformations (diffeomorphism).~\cite{Witten88, Atiyah88} This is because their physical properties do not depend on local structures of systems, and depend only on topological properties of geometric manifolds. A fairly complete class of two-dimensional TQFT-based spin systems with non-chiral topological order has been proposed by Levin and Wen where condensation of highly-fluctuating extended objects, called ``string-nets'', are found to be responsible for emergence of topological order.~\cite{Levin05}

Yet, in some cases, quantum spin liquids may exhibit topological order that is beyond description of TQFT. For example, in three spatial dimensions, the Cubic code, recently proposed by Haah~\cite{Haah11}, possesses topological order with stability against local perturbations, but are completely different from conventional topological spin liquids. For one thing, the number of degenerate ground states is exponential in the linear length of the lattice. Furthermore, unlike string-net condensates, the model is free from string-like extended objects, and the mobility of quasi-particle excitations is highly constrained via some algebraic rules. The discovery of the Cubic code and relevant models~\cite{Kim12, Castelnovo12} clearly indicates that classification of topological phases via TQFT is incomplete; TQFT is just a subset of some universal theory of topological order which is yet to be found. The necessary first step is to find a family of topological spin liquids that are beyond TQFT.

The goal of this paper is to present a large class of exactly solvable topological spin liquids on a three-dimensional lattice which possess exotic topological order beyond TQFT. Instead of string-like (one-dimensional) or membrane-like (two-dimensional) objects with continuous geometries, ground states are condensation of extended objects with non-integer dimensionality, namely \emph{fractal objects}. In this paper, we discuss physical properties of such \emph{quantum fractal liquids}.

Emergence of fractal objects in correlated spin systems is not a completely new idea. Newman and Moore proposed a toy model of two-dimensional \emph{classical} spin liquid with a large number of degenerate ground states whose spin configurations resemble the Sierpinski triangle.~\cite{Newman99} By generalizing their construction, we proved that a family of such fractal systems, refereed to as \emph{classical fractal liquids} in this paper, saturates a theoretical limit on classical information storage capacity of local Hamiltonians with mass gap.~\cite{Beni11b} As demonstrated in this paper, ground states of classical fractal liquids do not have continuous scale symmetries, but have discrete scale symmetries only, exhibiting \emph{limit cycle} behaviors under real-space RG transformations. Such exotic features of classical fractal liquids indicate a possibility of novel quantum phases beyond field theory with continuous scale invariance. Quantum fractal liquids can be viewed as natural generalization of classical fractal liquids to a quantum setting, and may potentially saturate quantum information storage capacity.

The paper is organized as follows. In section~\ref{sec:symmetry}, we first present a physical picture of quantum fractal liquids by reviewing how condensation of extended objects emerge in topological spin liquids. This section serves as non-technical summary of the paper. We then present a general framework to construct a family of classical fractal liquids in section~\ref{sec:CFL}. In section~\ref{sec:RG}, we demonstrate that ground states of classical fractal liquids correspond to limit cycles under real-space RG transformations. In section~\ref{sec:QFL}, we present a general framework to construct a family of three-dimensional quantum fractal liquids. In section~\ref{sec:no_string}, we discuss quasi-particle properties and look at several examples. In section~\ref{sec:conclusion}, we briefly discuss coding properties of quantum fractal liquids.

Some comments on the paper follow. We adopt the stability against local perturbations as the definition of topological order. By TQFT, we mean an axiomatic formulation by Atiyah which admits only a finite number of degenerate ground states.~\cite{Atiyah88, Nayak08} By topological spin liquids, we mean gapped spin systems without local symmetries, i.e. topologically ordered spin systems. Discussion on gapless quantum spin liquids is beyond the scope of this paper. Our construction of quantum fractal liquids is theoretically motivated, and its relevance to experimental realization may not be immediately clear. Some technical tools are borrowed from a recent work by Haah.~\cite{Haah12} 

\section{Topological spin liquid}\label{sec:symmetry}

In conventional topological spin liquids, extended objects with continuous geometries emerge from underlying gauge symmetries.~\cite{Read91, Wen91} In contrast, quantum fractal liquids are condensation of fractal objects with discrete geometries which emerge from certain algebraic symmetries. Geometric properties of extended objects can be characterized by topological classes of symmetry operators; fractal operators are associated with quantum fractal liquids. In this section, we present a physical picture of quantum fractal liquids. 

\subsection{Topological spin liquid and string-nets}\label{sec:symmetry1}

\begin{figure*}[htb!]
\centering
\includegraphics[width=0.70\linewidth]{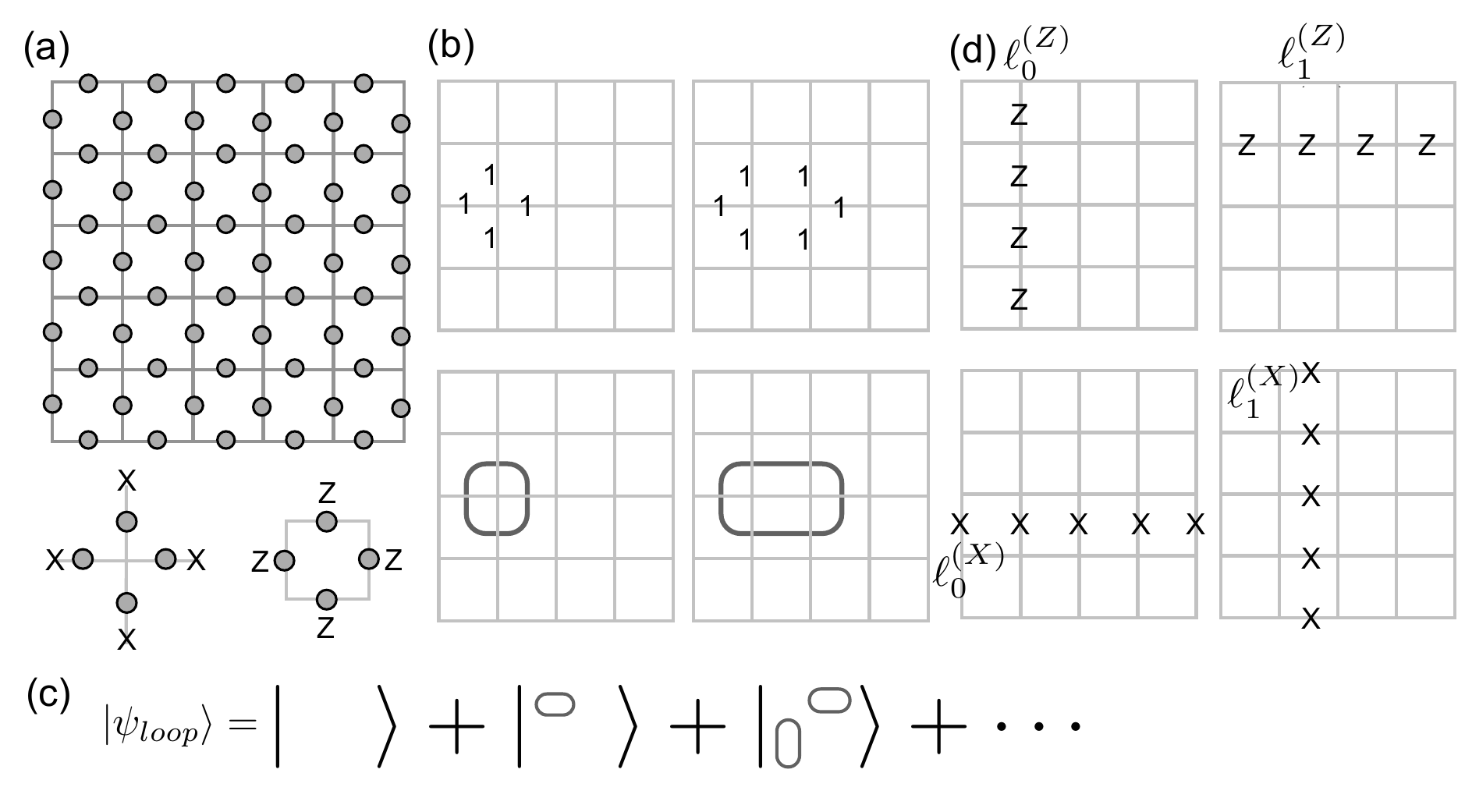}
\caption{$\mathbb{Z}_{2}$ spin liquid (the Toric code). (a) The Hamiltonian. (b) Loop states on a dual lattice. (c) Condensation of loops. (d) Logical operators.
} 
\label{fig_Toric}
\end{figure*}

We begin with the simplest string-net model, known as $\mathbb{Z}_{2}$ spin liquid (or the Toric code) (Fig~\ref{fig_Toric}(a)).~\cite{Kitaev03, Wen03} Consider a square lattice where qubits live on edges of the lattice with periodic boundary conditions. The Hamiltonian is 
\ba{
H = - \sum_{s} A_{s}- \sum_{p}B_{p},\quad A_{s} = \prod_{r\in s} X_{r},\quad 
B_{p}=\prod_{r\in p} Z_{r}
}
where $s$ represents a star and $p$ represents a plaquette. Pauli $X$ and $Z$ operators act on each qubit as follows: $Z|0\rangle=|0\rangle$, $Z|1\rangle=-|1\rangle$, $X|0\rangle=|1\rangle$  and $X|1\rangle=|0\rangle$. The model is exactly solvable as interaction terms $A_{s}$ and $B_{p}$ commute with each other, and ground states satisfy
\ba{
A_{s}|\psi\rangle = |\psi\rangle,\quad B_{p}|\psi\rangle =|\psi\rangle,\quad \forall s,p.
}
A ground state can be viewed as condensation of string-like extended objects. Consider a trivial product state $|0\rangle^{\otimes N}$ over the entire lattice ($N$ is the total number of qubits) and observe that $B_{p}|0\rangle^{\otimes N} = |0\rangle^{\otimes N}$. The following is a ground state:
\begin{align}
|\psi_{loop}\rangle = \prod_{s} (1+A_{s})  |0\rangle^{\otimes N}\label{eq:condensation}
\end{align}
since $A_{s}(1+A_{s})=1+A_{s}$. The normalization factor is omitted. It is a superposition of $A_{s_{1}}A_{s_{2}}A_{s_{3}}\cdots|0\rangle^{\otimes N}$. Since $A_{s}$ is a product of Pauli-$X$ operators, it flips qubits: $|0\rangle \leftrightarrow |1\rangle$. Then a term $A_{s}|0\rangle^{\otimes N}$ can be viewed as a state with one small loop on a dual lattice, and a term $A_{s_{1}}A_{s_{2}}|0\rangle^{\otimes N}$ with neighboring stars $s_{1}$ and $s_{2}$ is a state with a larger loop (Fig.~\ref{fig_Toric}(b)). In general, $A_{s_{1}}A_{s_{2}}A_{s_{3}}\cdots|0\rangle^{\otimes N}$ is a state with loops of various sizes and shapes. A ground state is a superposition of all the loop states (Fig.~\ref{fig_Toric}(c)):
\begin{align}
|\psi_{loop}\rangle = \sum_{\forall \gamma} |\gamma\rangle
\end{align}
where $\gamma$ represents an arbitrary contractable loop configuration. Therefore a ground state is condensation of \emph{fluctuating string-like objects} with $\mathbb{Z}_{2}$ gauge symmetry. 

One can construct a general quantum many-body system with several types of strings constrained by discrete gauge symmetry. Levin and Wen derived the most general form of wave-functions that are represented as condensation of string-like extended objects on a two-dimensional lattice by further assuming that wave-functions possess \emph{scale invariance} and correspond to fixed-points of RG transformations.~\cite{Levin05} Indeed, a ground state of $\mathbb{Z}_{2}$ spin liquid has scale invariance as it is a superposition of loops of all the different sizes and shapes. Note that scale invariance is required for systems described by TQFT since they must be invariant under continuous deformations. Yet, scale invariance is not a necessary condition for the presence of topological order. As we will see, quantum fractal liquids do not have full continuous scale symmetries. Instead, they have \emph{discrete scale symmetries} where systems are invariant only under a limited set of scale transformations and ground states correspond to limit cycles of RG transformations.

Geometric properties of extended-objects can be characterized by topological properties of global symmetry operators. Formally, symmetries of the Hamiltonian can be captured by unitary transformations that leave the Hamiltonian invariant:
\begin{align}
U^{\dagger}H U = H.
\end{align}
Interaction terms $A_{s}$ and $B_{p}$ are symmetry operators for $\mathbb{Z}_{2}$ spin liquid since $[A_{s},H]=[B_{p},H]=0$ where the ground state space is an invariant subspace under actions of interaction terms. There also exist \emph{non-trivial symmetry operators} which act non-trivially inside the ground state space (see Fig~\ref{fig_Toric}(d)):
\ba{
[H,\ell_{0}^{(Z)}] = [H,\ell_{1}^{(Z)}] = [H,\ell_{0}^{(X)}] = [H,\ell_{1}^{(X)}] = 0.\
}
with non-trivial winding on a torus. These symmetries are spontaneously broken in ground states.

Since non-trivial symmetry operators commute with the Hamiltonian, they do not change the energy. Yet, they cannot be written as products of $A_{s}$ or $B_{p}$ and transform degenerate ground states into each other. Recall that $|\psi_{loop}\rangle$ is a condensation of loops that can be shrunk into a vacuum under $\mathbb{Z}_{2}$ gauge symmetry. An application of $\ell_{0}^{(X)}$ to $|\psi_{loop}\rangle$ creates a non-trivial loop winding in the $\hat{x}$ direction. Similarly, $\ell_{1}^{(X)}|\psi_{loop}\rangle$ is condensation of loops with non-trivial winding in the $\hat{y}$ direction. Four degenerate ground states may be indexed by winding numbers as $|\tilde{\gamma_{x}}\rangle \otimes |\tilde{\gamma_{y}}\rangle$ with $\gamma_{x},\gamma_{y}=0,1$ where $\gamma_{x}$ and $\gamma_{y}$ represent the absence or presence of windings in the $\hat{x}$ and $\hat{y}$ directions respectively. Then non-trivial symmetry operators $\ell_{0}^{(X)}$ and $\ell_{1}^{(X)}$ act like Pauli-$X$ operators on a pair of \emph{logical qubits} $|\tilde{\gamma_{x}}\rangle \otimes |\tilde{\gamma_{y}}\rangle$. It is convenient to represent their commutation relation as follows
\ba{
\left\{
\begin{array}{cc}
\ell^{(Z)}_{0} ,& \ell^{(Z)}_{1}\\
\ell^{(X)}_{0} ,& \ell^{(X)}_{1}
\end{array}
\right\}
}
where operators in the same column anti-commute with each other while logical operators in different columns commute with each other. (This notation for commutation relations among logical operators will be used throughout the paper). Anti-commuting pairs of non-trivial symmetry operators can be viewed as logical Pauli-$Z$ and Pauli-$X$ acting on logical qubits $|\tilde{\gamma_{x}}\rangle \otimes |\tilde{\gamma_{y}}\rangle$. 

These non-trivial symmetry operators are often called \emph{logical operators} in quantum information community as they are utilized to rewrite encoded logical qubits in the ground state space. Note $\mathbb{Z}_{2}$ spin liquid is a good quantum error-correcting code as only global operators with non-trivial winding can change encoded logical qubits. This is an insight on stability against local perturbations in topological phases from quantum information perspective.~\cite{Kitaev03, Bravyi10b}

One can generalize $\mathbb{Z}_{2}$ spin liquid to higher-dimensional systems. For instance, the $D$-dimensional Toric code has  anti-commuting pairs of $m$-dimensional and $(D-m)$-dimensional logical operators:
\ba{
\mbox{$m$-dim} \ \leftrightarrow \ \mbox{$(D-m)$-dim}\qquad \mbox{$m$ : integer}
}
Its ground states can be viewed as condensation of $m$-dimensional extended-objects, or $(D-m)$-dimensional extended objects in a dual description. In general, for a quantum many-body system described by TQFT, extended objects (Wilson loops or Wilson surfaces) have continuous geometries with integer dimensionality. In fact, the dimensional duality of non-trivial symmetry operators is a consequence of the Poincar\'e duality for systems described by TQFT and can be derived from continuous deformability of logical operators.~\cite{Beni11}

\subsection{Emergence of fractal geometry}\label{sec:symmetry2}

In this subsection, we give a physical picture of quantum fractal liquids by reviewing how fractal geometries arise in classical spin systems. Consider a square lattice where $L\times L$ spins live on vertices. The Hamiltonian is
\ba{
&H = - \sum_{i=0}^{L-2}\sum_{j=0}^{L-1} \Pi_{ij}\\
&\Pi_{ij} = Z_{i,j}Z_{i+1,j}Z_{i+1,j+1} 
= \left(\begin{array}{cc}
Z_{i,j}, & Z_{i+1,j} \\
I, &Z_{i+1,j+1}
\end{array}\right)
} 
where $Z_{ij}$ acts on a spin at $(i,j)$ and we represented interaction terms graphically as a matrix. We denote spin values at $(i,j)$ for $i,j=0,\cdots,L-1$ as $s_{i,j}=0,1$. Ground states must satisfy $\Pi_{ij} \psi = \psi$: 
\begin{align}
s_{i,j} + s_{i+1,j} = s_{i+1,j+1}\quad \mbox{(mod $2$)} \label{eq:fractal}
\end{align}
for all $i$ and for $j=0,\cdots,L-2$. The following is a ground state:
\ba{
\begin{bmatrix}
1& 0 & 0 & 0 & 0 & 0 & 0 & 0 & \cdots \\
1& 1 & 0 & 0 & 0 & 0 & 0 & 0 & \cdots \\
1& 0 & 1 & 0 & 0 & 0 & 0 & 0 & \cdots \\
1& 1 & 1 & 1 & 0 & 0 & 0 & 0 & \cdots \\
1& 0 & 0 & 0 & 1 & 0 & 0 & 0 & \cdots \\
1& 1 & 0 & 0 & 1 & 1 & 0 & 0 & \cdots \\
1& 0 & 1 & 0 & 1 & 0 & 1 & 0 & \cdots \\
1& 1 & 1 & 1 & 1 & 1 & 1 & 1 & \cdots \\
\vdots& \vdots& \vdots& \vdots& \vdots& \vdots& \vdots& \vdots & \ddots
\end{bmatrix}
}
where the upper-left corner corresponds to $s_{0,0}$ and the lower-right corner corresponds to $s_{L-1,L-1}$. Configuration of sites with spin value $1$ forms the \emph{Sierpinski triangle} (Fig.~\ref{fig_fractal}(a)). It is interesting to observe that translation symmetries are spontaneously broken in a ground state. 

The model has a large number of degenerate ground states. Let us pick up arbitrary spin values on the first row $\vec{s}=(s_{0,0}, s_{1,0},s_{2,0},\cdots,s_{L-1,0})$. Then spin values on other rows are determined by Eq.~(\ref{eq:fractal}). Since there are $2^{L}$ possible choices for $\vec{s}$, there are $2^{L}$ degenerate ground states. It is convenient to view the model as a time-evolution of one-dimensional cellular automaton where spin values on lower rows are computed via an update rule in Eq.~(\ref{eq:fractal}). It is well known that one-dimensional cellular automata with linear update rules generate a variety of fractal geometries (see Ref.~\onlinecite{Wolfram_Text} for a review). One may consider a general class of classical spin models with fractal ground states by designing interaction terms which imitate update rules of one-dimensional cellular automata. 

The Sierpinski triangle model has liquid-like order, but is different from conventional classical spin liquids such as anti-ferromagnetic Ising models on geometrically frustrated lattices.~\cite{Ramirez94, Harris97, Moessner98, Henley05, Castelnovo08b} Due to unconventional three-body interactions, the model does not have magnetic order at any temperature including $T=0$. A zero temperature thermodynamic entropy is large, but not extensive: $S=\sqrt{N}$. Weather the model may select an ordered ground state via order by disorder mechanism is not known. In this paper, we refer to a family of the Sierpinski triangle model as \emph{classical fractal liquids} despite technical subtleties mentioned above.

The model does not have power law decay of two-point correlation functions as observed in conventional classical spin liquids. Instead, it has an oscillatory power-law behavior with imaginary scaling dimensions, exhibiting \emph{discrete scale symmetries}. Consider the following three-point correlation function:
\begin{align}
C(r) = \langle Z_{i,j}Z_{i+r,j}Z_{i+r,j+r} \rangle.
\end{align}
In the ground space manifold, $C(r)$ is
\ba{
C(r) &= 1, \quad r=2^{m} \\
     &= 0, \quad r\not=2^{m}
}
with oscillatory behaviors in $\log r$, instead of $r$. This is because $Z_{i,j}Z_{i+r,j}Z_{i+r,j+r}$ is a product of interaction terms only if $r=2^{m}$ (see Eq.~(\ref{eq:scale}) too). At finite temperature, the three-point correlation function reads
\ba{
C(r) = (1-2p)^{r^{\log 3 / \log2}}, \quad r={2^{m}}
}
with $p = e^{-\beta}/(e^{-\beta}+e^{\beta})$ where the exponent depends on the fractal dimension. The correlation function for all $r$ may be written as
\ba{
C(r) \propto \exp( - \mbox{const} \cdot r^{\log 3 / \log2})\cdot \sum_{k=-\infty}^{\infty} r^{i\frac{2\pi k}{\log 2}}
}
where oscillatory behaviors in $\log r$ are represented by power law. Note imaginary scaling dimension is characteristic of systems with discrete scale symmetries as pointed out by Wilson.~\cite{Wilson71} 

\begin{figure}[htb!]
\centering
\includegraphics[width=0.70\linewidth]{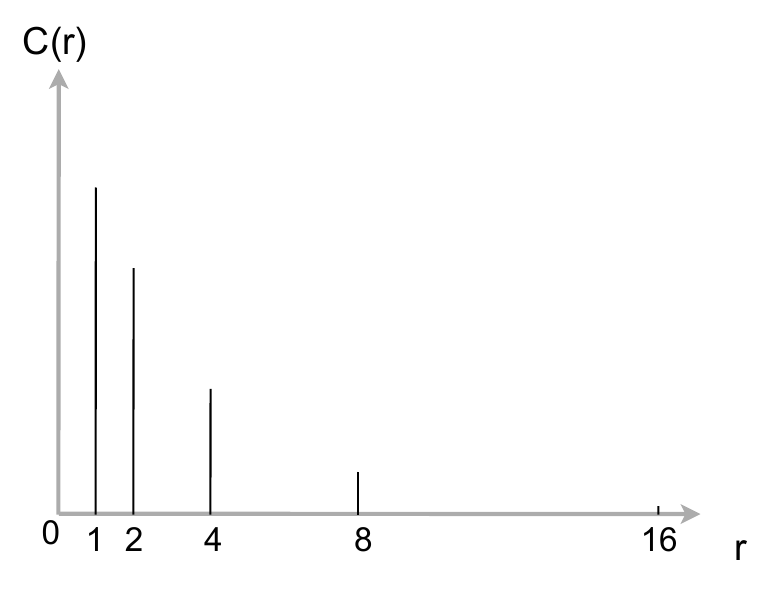}
\caption{Discrete scale symmetries and imaginary scaling dimensions in three-point correlation function.
} 
\label{fig_decay}
\end{figure}

Geometric properties of degenerate ground states in the Sierpinski triangle model can be captured by geometric shapes of logical operators:
\ba{\ell = 
\begin{bmatrix}
Z& I & I & I &  \cdots \\
I& I & I & I &  \cdots \\
I& I & I & I &  \cdots \\
I& I & I & I &  \cdots \\
\vdots& \vdots& \vdots& \vdots&  \ddots
\end{bmatrix},\quad r=
\begin{bmatrix}
X& I & I & I &  \cdots \\
X& X & I & I &  \cdots \\
X& I & X & I &  \cdots \\
X& X & X & X &  \cdots \\
\vdots& \vdots& \vdots& \vdots&  \ddots
\end{bmatrix}.
}
So, it has the following pairs of logical operators:
\ba{
\mbox{$0$-dim} \qquad \leftrightarrow \qquad\mbox{$\frac{\log 3}{\log 2}$-dim}.
}

While the model has a fractal logical operator, a ground state does not have any quantum fluctuation, and is not topologically ordered since its partner is a trivial logical operator with zero-dimensional geometry. To have topological order, both logical operators must have topologically non-trivial geometries (i.e. they must be finite-dimensional). In the reminder of the paper, we present a large class of quantum spin systems which has pairs of anti-commuting fractal logical operators:
\ba{
\mbox{fractal-dim} \qquad \leftrightarrow \qquad\mbox{fractal-dim}
}

\subsection{Topological phase transition}\label{sec:symmetry3}

A quantum many-body system with topological order can be viewed as condensation of extended objects with a variety of geometric shapes. It is natural to expect that two ground states with different types of extended objects belong to different topological phases. In this subsection, we make this intuition more precise by arguing that two spin systems with topologically different classes of logical operators are separated by quantum phase transitions.

Quantum phases are characterized by long-range entanglement of a many-body quantum system with mass gap at zero temperature. Ground states in different quantum phases cannot be connected continuously at the thermodynamic limit. Let us consider two ground states $|\psi_{A}\rangle$ and $|\psi_{B}\rangle$ of two different gapped Hamiltonian $H_{A}$ and $H_{B}$ and ask if they are separated by quantum phase transitions (non-analytic changes of ground state properties). Two ground states $|\psi_{A}\rangle$ and $|\psi_{B}\rangle$ are said to be in different quantum phases when there always exist quantum phase transitions between $H_{A}$ and $H_{B}$ regardless of paths between $H_{A}$ and $H_{B}$. Conversely, if there exists a continuous change from $H_{A}$ to $H_{B}$ without crossing quantum phase transitions, two ground states $|\psi_{A}\rangle$ and $|\psi_{B}\rangle$ are in the same quantum phase.

\begin{figure*}[htb!]
\centering
\includegraphics[width=0.85\linewidth]{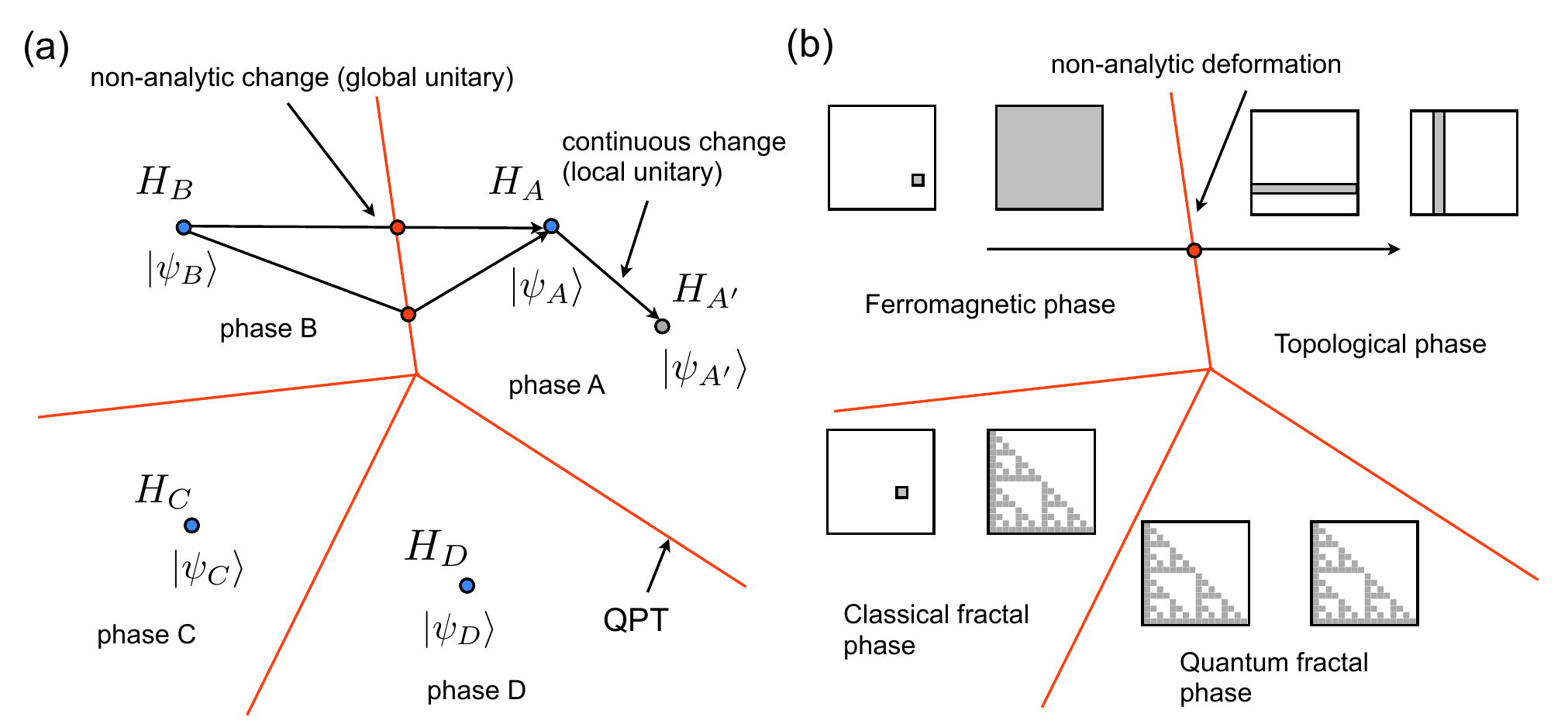}
\caption{(Color online) (a) Continuous deformability of ground states. (b) Continuous deformability of logical operators.
} 
\label{phase_diagram}
\end{figure*}

An equivalent, but more convenient way of classifying quantum phases uses local unitary transformations. Two ground states $|\psi_{A}\rangle$ and $|\psi_{B}\rangle$ are considered to be in the same quantum phase when there exists some local unitary transformation connecting $|\psi_{A}\rangle$ and $|\psi_{B}\rangle$. By local unitary transformations, we mean transformations generated by a set of geometrically local quantum operations, applied for a finite duration. On the other hand, when there is no local unitary transformation connecting $|\psi_{A}\rangle$ and $|\psi_{B}\rangle$, they are in different quantum phases.
Only global unitary transformations can change long-range entanglement of ground states. 

These two classification principles of quantum phases are equivalent under appropriate assumptions. If two gapped Hamiltonians $H_{A}$ and $H_{B}$ can be transformed into each other continuously without closing the energy gap, correlation lengths of ground states remain finite, and ground states at each stage of transformation can be approximated via some quasi-local unitary transformations applied to original ground states.~\cite{Bravyi10b} Conversely, if $|\psi_{A}\rangle$ and $|\psi_{B}\rangle$ are connected by some local unitary transformations, one can always continuously transform $H_{A}$ into $H_{B}$. 

The classification of quantum phases, based on continuous deformability of ground state wave-functions, reminds us of the study of \emph{topology} in mathematics, which aims to classify geometric shapes of object based on continuous deformability. Roughly speaking, two objects are considered to be equivalent when they can be transformed into each other via continuous deformations (diffeomorphism). Yet, if one cannot continuously deform an object to the other, they are considered to be topologically different. The similarity between classifications of quantum phases, based on continuous deformability of wave functions, and classifications of geometric shapes, based on continuous deformability of geometric objects, allows us to use the notion of topology in classifying quantum phases. Indeed, the following relation holds: 
\ba{
&\mbox{Logical operators are topologically different.} \\ &\Rightarrow \ \mbox{Two systems belong to different quantum phases.}
}
The argument roughly goes as follows.~\cite{Hastings05, Beni10b} Consider two systems with topologically distinct logical operators $\ell$ and $\ell'$. Let us suppose that they belong to the same quantum phase. Then, there must be some local unitary transformation $U$ such that $U\ell U^{\dagger}=\ell'$. Yet, this is not possible since local unitary transformation can change geometric shapes of logical operators only continuously at the thermodynamic limit. Therefore, models with topologically different types of logical operators belong to different quantum phases and are always separated by quantum phase transitions. Note that there is no local unitary that transforms a string-like logical operator to a fractal logical operator, and thus, fractal models are different from conventional topologically ordered systems. ``$\Leftarrow$'' of the above relation is proven only for stabilizer Hamiltonians in two-dimensions.~\cite{Beni10b} 

In summary, we expect that there will be four classes of quantum phases arising in gapped spin systems. 
\begin{enumerate}[(a)]
\item Ferromagnetic phase: $0$-dim $\leftrightarrow$ $D$-dim. 
\item Classical fractal phase: $0$-dim $\leftrightarrow$ fractal-dim. 
\item Topological phase: $m$-dim $\leftrightarrow$ $D-m$-dim ($m>0$). 
\item Quantum fractal phase: fractal-dim $\leftrightarrow$ fractal-dim.
\end{enumerate}

\section{Classical fractal liquid}\label{sec:CFL}

\subsection{Fractal and algebraic symmetry}\label{sec:CFL1}
 
\begin{figure*}[htb!]
\centering
\includegraphics[width=0.9\linewidth]{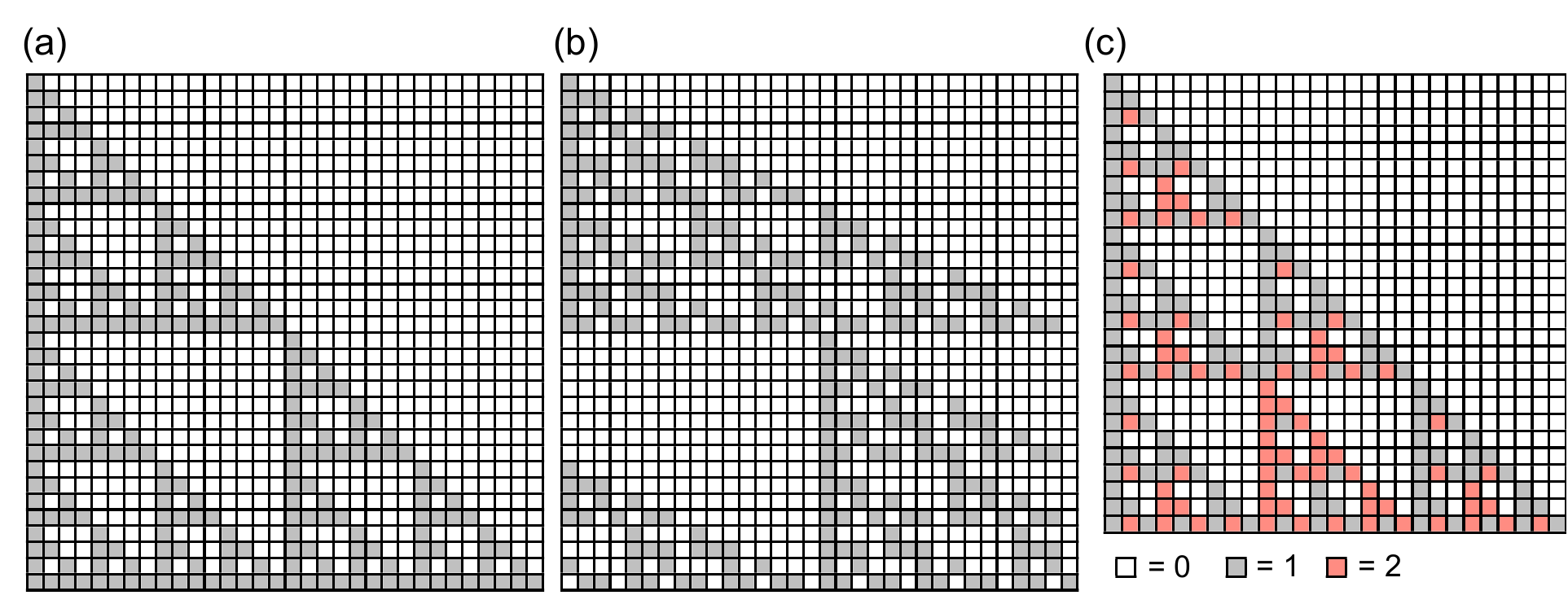}
\caption{(Color online) (a) The Sierpinski triangle from $f=1+x$ over $\mathbb{F}_{2}$. (b) The Fibonacci model from $f=1+x+x^{2}$ over $\mathbb{F}_{2}$. (c) The generalized Sierpinski triangle from $f=1+x$ over $\mathbb{F}_{3}$. 
} 
\label{fig_fractal}
\end{figure*}

In this section, we construct a family of classical fractal liquids. We begin with polynomial representation of the Sierpinski triangle (Fig.~\ref{fig_fractal}(a)). Consider a polynomial $f=1+x$ over $\mathbb{F}_{2}$ and its powers:
\ba{
f^{0}&=1\\
f^{1}&=1+x\\
f^{2}&=1+x^{2}\\
f^{3}&=1+x+x^{2}+x^{3}\\
f^{4}&=1 + x^{4}\\
f^{5}&=1+x+x^{4}+x^{5}
}
where coefficients are computed modulo $2$. More graphically, one has
\ba{
\begin{bmatrix}
f^{0}\\
f^{1}\\
f^{2}\\
f^{3}\\
f^{4}\\
f^{5}
\end{bmatrix}
=
\begin{bmatrix}
1 & & & & &\\
1 &x& & & &\\
1 & &x^{2}&&&\\
1 & x & x^{2} & x^{3}& & \\
1 & & & & x^{4}&\\
1 & x & & & x^{4}& x^{5}
\end{bmatrix}
}
where the Sierpinski triangle emerges in a geometric pattern of non-zero coefficients in $f^{j}$. 

The entire Sierpinski triangle can be represented as a single polynomial with $x$ and $y$:
\ba{
\tb{f}(x,y) = 1 + fy + f^{2}y^{2} + f^{3}y^{3} + \cdots
}
where $j$-th row is indexed by $y^{j}$. More graphically, one has
\ba{
\tb{f}(x,y)  =
\begin{bmatrix}
1 & & & & &\\
y &xy& & & &\\
y^{2} & &x^{2}y^{2}&&&\\
y^{3} & xy^{3} & x^{2}y^{3} & x^{3}y^{3}& & \\
y^{4} & & & & x^{4}y^{4}&\\
y^{5} & xy^{5} & & & x^{4}y^{5}& x^{5}y^{5}
\end{bmatrix}
}
where non-zero coefficients of $x^{i}y^{j}$ correspond to filled elements of the Sierpinski triangle at $(i,j)$.

Another interesting example of fractal geometries is generated by $f = 1+ x + x^{2}$ over $\mathbb{F}_{2}$:
\ba{
\begin{bmatrix}
f^{0}\\
f^{1}\\
f^{2}\\
f^{3}\\
f^{4}
\end{bmatrix}
=
\begin{bmatrix}
1 & & & & & & & & \\
1 &x& x^{2}& & & & & &\\
1 & &x^{2}& & x^{4} & & & &\\
1 & x &  & x^{3} & &x^{5} &x^{6} & & \\
1 & & & & x^{4} & & & & x^{8}
\end{bmatrix}.
}
Again, the entire fractal geometry can be represented as $\tb{f}(x,y)  = 1 + fy + f^{2}y^{2} + f^{3}y^{3} + \cdots$. The model is often called the Fibonacci model since its fractal dimension is given by $\frac{\log 1 + \sqrt{5}}{\log 2}$ (Fig.~\ref{fig_fractal}(b)). These constructions can be generalized to polynomials over $\mathbb{F}_{p}$ ($p>2$) with an arbitrary prime $p$. For instance, $f=1+x$ over $\mathbb{F}_{3}$ ($p=3$) leads to a generalization of the Sierpinski triangle for three-dimensional spins (Fig.~\ref{fig_fractal}(c)):
\ba{
\begin{bmatrix}
f^{0}\\
f^{1}\\
f^{2}\\
f^{3}\\
f^{4}\\
f^{5}
\end{bmatrix}
=
\begin{bmatrix}
1 & & & & &\\
1 &x& & & &\\
1 & 2x &x^{2}&&&\\
1 &   &    & x^{3} && \\
1 & x & & x^{3} & x^{4}& \\
1 & 2x & x^{2}& x^{3} & 2x^{4}& x^{5} 
\end{bmatrix}
}

The self-similarity in fractal geometries arises from \emph{discrete scale symmetries} of generating polynomials. Consider an arbitrary polynomial $f$ over $\mathbb{F}_{p}$ with prime $p$:
\begin{align}
f = c_{0} + c_{1} x + c_{2} x^{2} + c_{3} x^{3} + \cdots
\end{align}
where $c_{j}=0,\cdots,p-1$. Then its $p$-th power is
\begin{align}
f^{p} = c_{0} + c_{1} x^{p} + c_{2} x^{2p} + c_{3} x^{3p} + \cdots. \label{eq:scale}
\end{align}
For instance, with $f=1+2x+x^{2}$ over $\mathbb{F}_{3}$, one finds
\ba{
f = 1+ 2x + x^{2}, \ 
f^{3}=1+2x^{3}+x^{6}, \ 
f^{9}=1+2x^{9}+x^{18}
}
So, generated fractal geometry $\tb{f}(x,y)  = 1 + fy + f^{2}y^{2} + f^{3}y^{3} + \cdots$ has a self-similarity where the same pattern appear repeatedly at different length scales.

Fractal geometries do not possess gauge symmetries since growth of filled elements violates charge conservation where a single element may evolve into multiple elements of the same type in the $\hat{y}$ direction. This is in strong contrast with the fact that continuous geometries often have physical interpretations based on conservation laws associated with underlying gauge symmetries as in the case of TQFT.~\cite{Levin05} Charge conservation in scale invariant spin models originate from group theoretical constraints imposed on the parent Hamiltonian. Fractal geometries obey a more general form of symmetries, which are referred to as \emph{algebraic symmetries} in this paper, due to a possible relation to theory of algebraic geometry which concerns geometric structures of solutions of polynomial equations. 

\subsection{Polynomial representation of Pauli operators}\label{sec:CFL2}

To construct parent Hamiltonians of classical fractal liquids, it is convenient to represent interaction terms by polynomials too. Note this is a standard technique in classical coding theory.~\cite{Code_text} Consider a polynomial $f$ over $\mathbb{F}_{2}$:
\begin{align}
f = \sum_{j=-\infty}^{\infty} c_{j}x^{j}, \quad c_{j}=0,1.
\end{align}
We define the corresponding Pauli operators as follows
\begin{align}
Z(f) = \prod_{j=-\infty}^{\infty} Z_{j}^{c_{j}}, \quad X(f) = \prod_{j=-\infty}^{\infty} X_{j}^{c_{j}}
\end{align}
where $Z_{j}$ and $X_{j}$ are Pauli operators acting on $j$-th qubit. So, a polynomial $f$ encodes positions of qubits where Pauli operators $Z_{j}$ or $X_{j}$ may act. For instance, $f=1+x+x^{2}$ and $Z(f) = Z_{0}Z_{1}Z_{2}$. 

The polynomial representation of Pauli operators is particularly useful for studying spin systems with translation symmetries since translations can be concisely represented in terms of polynomials. For instance, consider a Pauli operator $Z(f)=Z_{0}Z_{1}Z_{2}$ for $f=1+x+x^{2}$. Then, its translation in the $\hat{x}_{+}$ direction is given by $Z_{1}Z_{2}Z_{3}$, whose polynomial representation is $Z(xf)$: 
\ba{
f = 1 + x + x^{2}      \quad& \rightarrow \quad  xf = x + x^{2} + x^{3}\\
Z(f) = Z_{0}Z_{1}Z_{2}\quad & \rightarrow \quad  Z(xf) = Z_{1}Z_{2}Z_{3}.
}
In general, $Z(xf)$ is a translation of $Z(f)$ in the $\hat{x}_{+}$ direction. Similarly, a translation in the $\hat{x}_{-}$ direction is given by $Z(x^{-1}f)$. One may generalize this formalism to higher-dimensional systems by adding extra variables $y,z,\cdots$.

To gain more insights, let us represent one-dimensional ferromagnet by polynomials over $\mathbb{F}_{2}$:
\ba{
H = - \sum_{j} Z(x^{j}(1+x))
}
where $Z(x^{j}(1+x)) = Z_{j}Z_{j+1}$. The Sierpinski triangle model, introduced in the previous section, is 
\ba{
H = - \sum_{ij} Z(x^{i}y^{j}(1+x+xy))
}
where interaction terms are translations of $Z(1+x+xy)$. In general, one may consider a classical translation symmetric Hamiltonian
\begin{align}
H = - \sum_{i,j,\cdots} Z(x^{i}y^{j}\cdots\alpha)
\end{align}
with an arbitrary polynomial $\alpha(x,y,\cdots)$. Ground states obey
\begin{align}
Z(x^{i}y^{j}\cdots\alpha)|\psi\rangle = |\psi\rangle,\quad \forall i,j,\cdots.
\end{align} 

The polynomial representation of Pauli operators becomes particularly powerful in analyzing commutation relations between Pauli operators $Z(f)$ and $X(g)$. Since we are interested in translation symmetric systems, we want to obtain commutation relations between $Z(f)$ and translations of $X(g)$. Let us imagine that we check commutation relations between $Z(f)$ and $X(x^{j}g)$ for all $j$ and assign integers $d_{j}=0,1$ as follows 
\ba{
&d_{j}=0 \qquad \mbox{for}\quad [Z(f),X(x^{j}g)]=0 \\
&d_{j}=1 \qquad \mbox{for}\quad \{Z(f),X(x^{j}g)\}=0.
}
Based on $d_{j}$, we define the \emph{commutation polynomial} $P(f,g)$ as follows:
\begin{align}
P(f,g) = \sum_{j}d_{j}x^{j}
\end{align}
such that
\begin{align}
Z(f)X(x^{j}g) = (-1)^{d_{j}}X(x^{j}g)Z(f).
\end{align}
Thus, the commutation polynomial $P(f,g)$ is a collection of commutation relations between $Z(f)$ and $X(x^{j}g)$. For instance, with $f=1+x+x^{2}$ and $g = 1+x$, $Z(f)$ anti-commutes only with $X(x^{-1}g)$ and $X(x^{2}g)$. So, the commutation polynomial is $P(f,g) = x^{-1} + x^{2}$.

The commutation polynomial $P(f,g)$ can be concisely written by introducing the notion of \emph{dual}:
\begin{align}
f = \sum_{j=-\infty}^{\infty} c_{j}x^{j}\quad \rightarrow \quad \bar{f} = \sum_{j=-\infty}^{\infty} c_{j} x^{-j}
\end{align}
where the dual $\bar{f}$ is obtained by taking $x \rightarrow x^{-1}$. Then, the commutation polynomial is given by the \emph{convolution}
\begin{align}
P(f,g) = f \bar{g} \label{eq:commutation}
\end{align}
For instance, one has $f\bar{g}=(1+x+x^{2})(1+x^{-1})= x^{-1} + 2+ 2x + x^{2} = x^{-1} +x^{2}$ for the above example. The proof of Eq.~(\ref{eq:commutation}) is straightforward by explicit calculation. Generalization to polynomials over $\mathbb{F}_{p}$ is also straightforward by using generalized Pauli matrices for $\mathbb{Z}_{p}$. 

Periodic boundary conditions can be introduced by imposing $x^{L}=1$. Below, reversibility of polynomial $f(x)$ becomes important. Let $f=\sum_{j}c_{j}x^{j}$ over $\mathbb{F}_{p}$. When $L=p^{m}$, one has 
\ba{
f^{L}=\sum_{j}c_{j}x^{Lj}=\sum_{j}c_{j} = f(1)
}
due to discrete scale symmetries and $x^{L}=1$. A polynomial $f$ is reversible if and only if $f(1)\not=0$. We say that $f$ is \emph{properly normalized} when $f(1)=1$ so that $f^{L}=1$.

\subsection{Classical fractal liquid}\label{sec:CFL3}

We present general construction of classical fractal liquids. Consider a two-dimensional square lattice with $L\times L$ spins ($L=2^{m}$) over $\mathbb{F}_{2}$. The Hamiltonian is 
\begin{align}
H = - \sum_{ij}Z(x^{i}y^{j} \bar{\alpha}),\quad \alpha = 1 - f(x) y
\end{align} 
where $f(x)$ is an arbitrary polynomial over $\mathbb{F}_{2}$ with $x$ only. We put periodic boundary conditions both in the $\hat{x}$ and $\hat{y}$ directions, and assume that $f(x)$ is reversible and properly normalized. 

In finding ground states of the Hamiltonian, it is convenient to find its logical operators. $Z$-type logical operators are trivial single Pauli operators $\ell^{(Z)}_{j} = Z_{j0}=Z(x^{j})$ for $j=0,\cdots,L-1$ while $X$-type logical operators have fractal geometries:
\ba{
\ell^{(X)}_{j} = X(x^{j}\tb{f}(x,y)),\quad \tb{f}(x,y) = 1 + fy + \cdots + (fy)^{L-1}.
}
One can see that $\ell^{(X)}_{j}$ commute with all the stabilizer generators since the commutation polynomial between $Z(1+\bar{f}\bar{y})$ and $X(\tb{f}(x,y))$ is  
\ba{
(1 - \bar{f}\bar{y})(1 + \bar{f}\bar{y} + \cdots + (\bar{f}\bar{y})^{L-1})=0
}
when $f(x)$ is properly normalized. We list all the logical operators as follows:
\ba{
\left\{
\begin{array}{ccc}
\ell^{(X)}_{0},\cdots, \ell^{(X)}_{L-1} \\
\ell^{(Z)}_{0},\cdots, \ell^{(Z)}_{L-1}
\end{array}
\right\}.
}
So, there are $k=L$ logical bits in total.

By using X-type logical operators $\ell^{(X)}_{j}$, one can find all the ground states of a classical fractal liquid. Let us denote spin values at $(i,j)$ as $s_{ij}=0,1$ for $i,j=0,\cdots,L-1$, and represent a ground state $\psi$ as
\begin{align}
\psi = \sum_{ij} s_{ij}x^{i}y^{j}.
\end{align}
Since the Hamiltonian consists only of $Z$-type Pauli operators, $\psi=0$ with $s_{ij}=0$ is a ground state of the Hamiltonian. (Recall $Z|0\rangle = |0\rangle$ and $Z|1\rangle = -|1\rangle$ in our notation). To find another ground state, one applies $\ell^{(X)}_{0}$ to $\psi = 0$ and obtains a fractal ground state:
\begin{align}
\psi(1) =1 + fy + \cdots + (fy)^{L-1}= \tb{f}(x,y)
\end{align}
One can find all the other ground states by applying fractal logical operators $\ell^{(X)}_{j}$. There are $2^{L}$ degenerate ground states, represented by
\begin{align}
\psi(\gamma) = \gamma(x) \tb{f}(x,y)
\end{align}
where $\gamma(x)$ is an arbitrary polynomial with $x$ only. Noting $\dim \gamma = L$, one finds $k=L$. 

Classical fractal liquids discussed so far are based on \emph{first-order} cellular automata whose present states at $t=\tau$ depend on states at $t=\tau-1$. In higher-order cellular automata, the present states at $t=\tau$ may depend on states at $t=\tau-q,\cdots,\tau-1$ for $q>1$. One can construct classical fractal liquids based on higher-order cellular automata by taking
\ba{
\alpha = 1 + f_{1}(x)y + f_{2}(x)y^{2} + \cdots + f_{q}(x)y^{q}.
}
However, it is generally difficult to write down spin configurations of higher-order classical fractal liquids explicitly. 

Since the model does not have gauge symmetries, its quasi-particle excitations violate charge conservation and propagate according to algebraic symmetries imposed by generating polynomial $f(x)$. Recall that ground states $\psi$ satisfy $Z(x^{i}y^{j}\bar{\alpha})\psi=\psi$ for all $i,j$, and quasi-particle excitations may be viewed as violations of these algebraic constraints. It is convenient to represent positions of excitations by an \emph{excitation polynomial}:
\ba{
E(x,y)=\sum_{i,j,\ell} c_{ij}x^{i}y^{j}
}
where an excited state $\psi'$ has
\ba{
&c_{ij}=0 \quad \mbox{for}\ \ Z(x^{i}y^{j}\bar{\alpha})\psi'= + \psi'\\
&c_{ij}=1 \quad \mbox{for}\ \ Z(x^{i}y^{j}\bar{\alpha})\psi'= - \psi'
}
such that a quasi-particle is present at $(i,j)$ if and only if $c_{ij}=1$. 

\begin{figure}[htb!]
\centering
\includegraphics[width=0.55\linewidth]{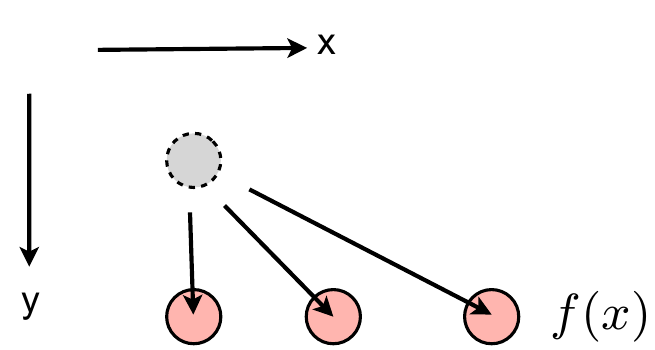}
\caption{(Color online) Propagation of quasi-particle excitations.
} 
\label{fig_excitation_C}
\end{figure}

Excitations in classical spin liquids are caused by Pauli-$X$ spin flips. Consider quasi-particle excitations caused by $X(e(x,y))$ where $e(x,y)$ are polynomials representing positions of spin flips. Since anti-commutations between $e(x,y)$ and $Z(x^{i}y^{j}\bar{\alpha})$ create quasi-particles at $(i,j)$, the excitation polynomial is 
\begin{align}
E(x,y)=e(x,y)\alpha.
\end{align}
For instance, if $X_{0,0}$ with $e=1$ is applied, one has multiple excitations $E(x,y)=\alpha$. Consider an isolated excitation at $(0,0)$. An application of $X_{0,0}$ makes it propagate in the $\hat{y}$ direction to multiple excitations represented by $f(x)y$ (Fig.~\ref{fig_excitation_C}). So, quasi-particle excitations propagate via applications of $f(x)$ like time-evolution of one-dimensional cellular automaton. Since the model does not have gauge symmetries, one cannot associate conserved charge to quasi-particle excitations. Indeed, a single quasi-particle may split into multiple quasi-particles of the same type. 

\section{Limit cycle under RG transformation}\label{sec:RG}

In this section, we demonstrate that ground states of classical fractal liquids correspond to limit cycles under real-space RG transformations. Treatment in this section can be applied to quantum fractal liquids too.

\subsection{Discrete scale symmetry}

Let us consider the Sierpinski triangle model ($f=1+x$) for $L=8$. A ground state is 
\ba{
\psi = \begin{bmatrix}
1 &  0 &  0 &  0 &  0 &  0 &  0 &  0 \\
1 &  1 &  0 &  0 &  0 &  0 &  0 &  0 \\
1 &  0 &  1 &  0 &  0 &  0 &  0 &  0 \\
1 &  1 &  1 &  1 &  0 &  0 &  0 &  0 \\
1 &  0 &  0 &  0 &  1 &  0 &  0 &  0 \\
1 &  1 &  0 &  0 &  1 &  1 &  0 &  0 \\
1 &  0 &  1 &  0 &  1 &  0 &  1 &  0 \\
1 &  1 &  1 &  1 &  1 &  1 &  1 &  1 
\end{bmatrix}.
}
RG transformations, denoted by $\mbox{RG}_{ij}$ ($i,j=0,1$), pick up spins at $(x,y)$ with $x=i$ (mod $2$) and $y=j$ (mod $2$), and throw away the rest:
\ba{
\mbox{RG}_{0,0}(\psi) = \begin{bmatrix}
1 &  0&  0&  0 \\
1 &  1&  0&  0 \\
1 &  0&  1&  0 \\
1 &  1&  1&  1
\end{bmatrix}\quad
\mbox{RG}_{0,1}(\psi) = \begin{bmatrix}
0 &  0&  0&  0 \\
0 &  0&  0&  0 \\
0 &  0&  0&  0 \\
0 &  0&  0&  0
\end{bmatrix}\\
 \mbox{RG}_{1,0}(\psi) = \begin{bmatrix}
1 &  0&  0&  0 \\
1 &  1&  0&  0 \\
1 &  0&  1&  0 \\
1 &  1&  1&  1
\end{bmatrix}\quad 
\mbox{RG}_{1,1}(\psi) = \begin{bmatrix}
1 &  0&  0&  0 \\
1 &  1&  0&  0 \\
1 &  0&  1&  0 \\
1 &  1&  1&  1
\end{bmatrix}
}
All the RG'ed states are ground states of the Hamiltonian for $L=4$. Let us look at another ground state
\ba{
\psi = \begin{bmatrix}
1 &  1&  0&  0&  0&  0&  0&  0 \\
1 &  0&  1&  0&  0&  0&  0&  0 \\
1 &  1&  1&  1&  0&  0&  0&  0 \\
1 &  0&  0&  0&  1&  0&  0&  0 \\
1 &  1&  0&  0&  1&  1&  0&  0 \\
1 &  0&  1&  0&  1&  0&  1&  0 \\
1 &  1&  1&  1&  1&  1&  1&  1 \\
0 &  0&  0&  0&  0&  0&  0&  0 
\end{bmatrix}
}
and its RG transformations:
\ba{
\mbox{RG}_{0,0}(\psi) = \begin{bmatrix}
1 &  0&  0&  0 \\
1 &  1&  0&  0 \\
1 &  0&  1&  0 \\
1 &  1&  1&  1
\end{bmatrix}\quad
\mbox{RG}_{0,1}(\psi) = \begin{bmatrix}
1 &  0&  0&  0 \\
1 &  1&  0&  0 \\
1 &  0&  1&  0 \\
1 &  1&  1&  1
\end{bmatrix}\\
 \mbox{RG}_{1,0}(\psi) = \begin{bmatrix}
1 &  1&  0&  0 \\
1 &  0&  1&  0 \\
1 &  1&  1&  1 \\
0 &  0&  0&  0
\end{bmatrix}\quad 
\mbox{RG}_{1,1}(\psi) = \begin{bmatrix}
0 &  0&  0&  0 \\
0 &  0&  0&  0 \\
0 &  0&  0&  0 \\
0 &  0&  0&  0
\end{bmatrix}
}
Again, RG'ed states are ground states of a smaller system. 

\begin{figure}[htb!]
\centering
\includegraphics[width=0.60\linewidth]{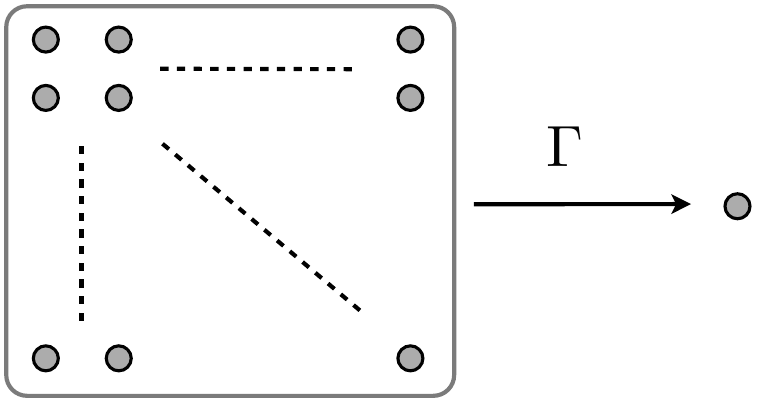}
\caption{RG transformation as a linear map $\Gamma$. 
} 
\label{fig_RG}
\end{figure}

In general, we consider the Sierpinski triangle model for $L=2^{m}$ and $\psi \in \mathcal{G}_{m}$ where $\mathcal{G}_{m}$ is a set of ground states. We view an arbitrary linear map $\Gamma$ from $2^{a}\times 2^{a}$ spins a single spin as an RG function (Fig.~\ref{fig_RG}):
\ba{
\Gamma : (\mathbb{F}_{2})^{\otimes 2^{2a}} \rightarrow \mathbb{F}_{2}
}
where $\Gamma$ maps a wavefunction for $L=2^{m}$ to a wavefunction for $L=2^{m-a}$. Then one has
\ba{
\Gamma(\psi) \in \mathcal{G}_{m-a}\quad \forall \psi\in \mathcal{G}_{m}
}
where $\mathcal{G}_{m-a}$ is a set of ground states for $L=2^{m-a}$. One can find RG transformations that are stable against small perturbations added to wave-functions so that it makes sense to discuss how wavefunctions flow under RG transformations. 

\begin{figure*}[htb!]
\centering
\includegraphics[width=0.95\linewidth]{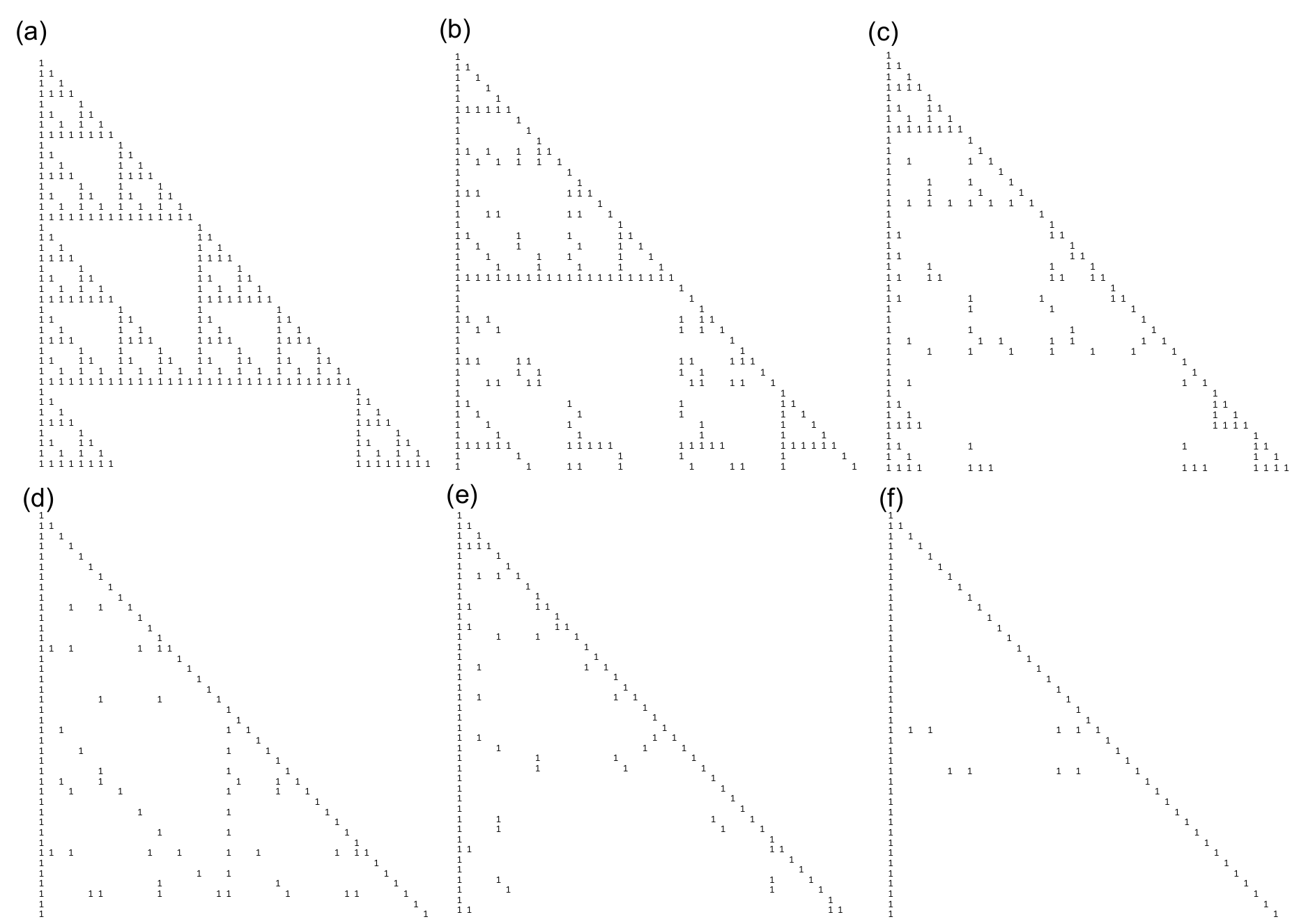}
\caption{Incommensurate RG of the Sierpinski triangle model. Spins at $(i,j)=(0,0)$ (mod $3$) are picked up. (a) The original. (b) $\lambda=3$. (c) $\lambda=9$. (d) $\lambda=27$. (e) $\lambda=81$. (f) $\lambda=243$. 
} 
\label{fig_incommensurate}
\end{figure*}

Ground states of the Sierpinski triangle model behave nicely under scale transformations by a factor of $2$ only. If one performs a similar RG transformation by an incommensurate factor, RG'ed states are not ground states of the Hamiltonian anymore and flow to something else. See Fig.~\ref{fig_incommensurate} for RG by factors of $\lambda=3^{m}$ where density of $1$ states decreases. So, the model has scale symmetries under some limited set of scale transformations. This is a striking contrast with the fact that a ferromagnet, a spin model with continuous scale symmetries, look always the same under any scale transformations. Note that among four RG'ed ground states, only two of them are independent. 

It turns out that presence of discrete scale symmetries is a general property of classical Hamiltonians with interaction terms $Z(\bar{\alpha})$ for an \emph{arbitrary polynomial} $\alpha$ over $\mathbb{F}_{p}$ in \emph{any spatial dimensions}. We denote RG functions as $\mbox{RG}_{ij}$ ($i,j=0,\cdots,p-1$) which pick up spins at $(x,y)$ where $x=i$ (mod $p$) and $y=j$ (mod $p$). When the Hamiltonian consists of $Z(\alpha)$, a ground state $\psi$ satisfies $Z(x^{i}y^{j}\cdots\alpha)\psi=\psi$ for all $i,j,\cdots$. So, one has $Z(\gamma \alpha)\psi=\psi$ for arbitrary polynomial $\gamma$. For $\gamma=\alpha^{p-1}$, one has $Z(\alpha^{p})\psi = \psi$. From Eq.~(\ref{eq:scale}), $Z(\alpha^{p})$ has supports only on sites $(i,j,\cdots)=(0,0,\cdots)$ (mod $p$). So, one has $Z(\alpha)\mbox{RG}_{00}(\psi)=\mbox{RG}_{00}(\psi)$, and $\mbox{RG}_{00}(\psi)$ is a ground state for a smaller system. Since $\Gamma$ can be represented as linear combination of $\mbox{RG}_{ij}$, $\Gamma(\psi)$ is also a ground state.

For simplicity of discussion, we concentrate on two-dimensional cases with $Z(\bar{\alpha})$ and $\alpha=1-f(x)y$. Let us represent a ground state as $\psi_{m}=\gamma(x) \tb{f}_{m}(x,y)$ where $L=p^{m}$ and $\tb{f}_{m}(x,y)$ is the polynomial representation of the fractal. Then, RG'ed states $\mbox{RG}_{ij}(\psi_{m})$ is always a ground state of the Hamiltonian for $L=p^{m-1}$. In particular, a polynomial $\gamma'(x)$ satisfying the following equation always exists:
\begin{align}
\mbox{RG}_{ij}(\gamma(x) \tb{f}_{m}) = \gamma'(x) \tb{f}_{m-1}.
\end{align}

Below, we look at several examples. For $f=1+x$ over $\mathbb{F}_{3}$, one has
\ba{
 \psi(1) =\begin{bmatrix}
1 &  0&  0&  0&  0&  0&  0&  0&  0\\
1 &  1&  0&  0&  0&  0&  0&  0&  0\\
1 &  2&  1&  0&  0&  0&  0&  0&  0\\
1 &  0&  0&  1&  0&  0&  0&  0&  0 \\
1 &  1&  0&  1&  1&  0&  0&  0&  0\\
1 &  2&  1&  1&  2&  1&  0&  0&  0 \\
1 &  0&  0&  2&  0&  0&  1&  0&  0 \\
1 &  1&  0&  2&  2&  0&  1&  1&  0 \\
1 &  2&  1&  2&  1&  2&  1&  2&  1  
\end{bmatrix}
}
and
\ba{
 \mbox{RG}_{0,0}(\psi(1)) =\begin{bmatrix}
1 &  0&  0\\
1 &  1&  0\\
1 &  2&  1
\end{bmatrix}\quad
 \mbox{RG}_{1,2}(\psi(1)) = \begin{bmatrix}
2 &  0&  0 \\
2 &  2&  0 \\
2 &  1&  2
\end{bmatrix}
}
For the Fibonacci model $f=1+x+x^{2}$ over $\mathbb{F}_{2}$, one has:
\ba{
\psi(1) = \begin{bmatrix}
1 &  0&  0&  0&  0&  0&  0&  0&  0& \cdots  \\
1 &  1&  1&  0&  0&  0&  0&  0&  0& \cdots  \\
1 &  0&  1&  0&  1&  0&  0&  0&  0& \cdots  \\
1 &  1&  0&  1&  0&  1&  1&  0&  0& \cdots  \\
1 &  0&  0&  0&  1&  0&  0&  0&  1& \cdots  \\
\vdots & \vdots &\vdots &\vdots &\vdots &\vdots &\vdots &\vdots &\vdots & \ddots
\end{bmatrix}.
}
RG'ed states are
\ba{
\mbox{RG}_{00}(\psi(1)) = \begin{bmatrix}
1 &  0&  0&  0&  0&  0&  0&  0&  0& \cdots  \\
1 &  1&  1&  0&  0&  0&  0&  0&  0& \cdots  \\
1 &  0&  1&  0&  1&  0&  0&  0&  0& \cdots  \\
1 &  1&  0&  1&  0&  1&  1&  0&  0& \cdots  \\
1 &  0&  0&  0&  1&  0&  0&  0&  1& \cdots  \\
\vdots & \vdots &\vdots &\vdots &\vdots &\vdots &\vdots &\vdots &\vdots & \ddots
\end{bmatrix}}
and
\ba{
\mbox{RG}_{01}(\psi(1)) = \begin{bmatrix}
1 &  1&  0&  0&  0&  0&  0&  0&  0& \cdots  \\
1 &  0&  0&  1&  0&  0&  0&  0&  0& \cdots  \\
1 &  1&  1&  1&  1&  1&  0&  0&  0& \cdots  \\
1 &  0&  1&  1&  1&  1&  0&  1&  0& \cdots  \\
1 &  1&  0&  0&  1&  1&  0&  0&  1& \cdots  \\
\vdots & \vdots &\vdots &\vdots &\vdots &\vdots &\vdots &\vdots &\vdots & \ddots
\end{bmatrix}
}

Discrete scale symmetries also arise at finite temperature as seen in distribution patterns of quasi-particle excitations. Consider an excited state with an excitation polynomial $E(x,y)$. An excitation energy $\Delta$ is given by the weight of excitation polynomial
\ba{
\Delta = 2 W\big(E(x,y)\big) 
}
where $W\big(E(x,y)\big)$ counts the number of non-zero coefficients in $E(x,y)$. Let $\mathcal{D}_{\Delta }$ be a set of excitation polynomials with an excitation energy $\Delta$
\ba{
\mathcal{D}_{\Delta } = \Big\{ E(x,y) : \Delta = 2 W\big(E(x,y)\big) \Big\}.
}
Note 
\ba{
E(x,y) \in  \mathcal{D}_{\Delta} \quad \Rightarrow \quad E(x,y)^{p} \in  \mathcal{D}_{\Delta}
}
where $E(x,y)^{p}$ is a dilation of $E(x,y)$ by $p$. So, an excitation set $\mathcal{D}_{\Delta}$ is invariant under dilation by factor of $p$, and excitation pattern at finite temperature have discrete scale symmetries.

In condensed matter physics, one often encounters phase transition models governed by fixed points which exhibit dynamical scaling 
\ba{
t \rightarrow \lambda^{z}t, \quad x\rightarrow\lambda x
}
where $z$ is called dynamical scaling exponent. In conformally invariant systems, one always finds $z=1$. Examples of the anisotropic scale invariance with $z=2$ at Lifshitz point often appears in condensed matter physics too. Classical fractal liquids correspond to cases with $z=0$ since excitation patterns have discrete scale symmetries for fixed energy although they are not at criticality. 

\subsection{Limit cycles} 

Discrete scale symmetries provide an useful algorithm to compute the fractal dimension of $\tb{f}(x,y)$. We illustrate the algorithm for the Fibonacci model: $f=1+x+x^{2}$ over $\mathbb{F}_{2}$. We denote a ground state with an initial condition $\gamma$ as $\psi(\gamma)$. Then, renormalization of ground states $\psi_{m}(1)$ and $\psi_{m}(1+x)$ gives the following ground states for $L=2^{m-1}$:
\ba{
&\mbox{RG}_{00}(\psi_{m}(1))=\psi_{m-1}(1)&\mbox{RG}_{00}(\psi_{m}(1+x))=\psi_{m-1}(1)\\
&\mbox{RG}_{10}(\psi_{m}(1))=\psi_{m-1}(0) &\mbox{RG}_{10}(\psi_{m}(1+x))=\psi_{m-1}(1)\\
&\mbox{RG}_{01}(\psi_{m}(1))=\psi_{m-1}(1+x)&\mbox{RG}_{01}(\psi_{m}(1+x))=\psi_{m-1}(1)\\
&\mbox{RG}_{11}(\psi_{m}(1))=\psi_{m-1}(1)&\mbox{RG}_{11}(\psi_{m}(1+x))=\psi_{m-1}(x)
}
Let us denote the weights of $\psi_{m}(1)$ and $\psi_{m}(1+x)$ as $A_{m}$ and $B_{m}$. Then, one has
\begin{align}
\left(
\begin{array}{c}
A_{m}\\
B_{m}
\end{array}
\right)
= \left(
\begin{array}{cc}
2 & 1 \\
4 & 0
\end{array}
\right)\left(\begin{array}{c}
A_{m-1}\\
B_{m-1}
\end{array}
\right).
\end{align}
This matrix has eigenvalues $1 \pm \sqrt{5}$, and thus, $A_{m}$ and $B_{m}$ scale as $O(L^{\frac{\log 1 + \sqrt{5}}{\log 2}})$ for large $L$. 

The above RG transformations concern classical fractal liquids on a finite lattice. If one performs RG transformations on an infinite lattice, $\mbox{RG}_{ij}(\psi)$ becomes a group operation where $\mbox{RG}_{ij}(\psi)$ is a linear map inside the ground state space. In the case of a ferromagnet, the RG functions are always trivial; $\mbox{RG}_{ij}(\psi)=\psi$ since $\psi$ is spatially uniform. Yet, for classical fractal liquids, $\mbox{RG}_{ij}(\psi)$ may be different from $\psi$ in general. 

This gives an interesting possibility of limit-cycle behaviors under RG transformations. Consider $f=1+x$ over $\mathbb{F}_{3}$. Let us apply an RG transformation for a ground state $\psi(1) = \tb{f}(x,y)$ where $\tb{f}(x,y)=1+fy + f^{2}y^{2}+\cdots$ is defined on a infinite lattice. Then, $\mbox{RG}_{12}(\psi)$ gives the following sequence
\ba{
\psi(1) \rightarrow \psi(2) \rightarrow \psi(1) \rightarrow \psi(2) \rightarrow \cdots
}
where a ground state $\psi(1)$ jumps to a different ground state $\psi(2)$, and the RG sequence exhibits a limit-cycle behavior. Next, for $f=1+x+x^{2}$ over $\mathbb{F}_{2}$, consider a ground state $\psi(1)=\tb{f}(x,y)$. Then, one has the following sequence under $\mbox{RG}_{01}$: 
\ba{
\psi(1) \rightarrow \psi(1+x) \rightarrow \psi(1) \rightarrow \psi (1+x) \rightarrow \cdots
}
which is also a limit cycle. 

\begin{figure}[htb!]
\centering
\includegraphics[width=0.9\linewidth]{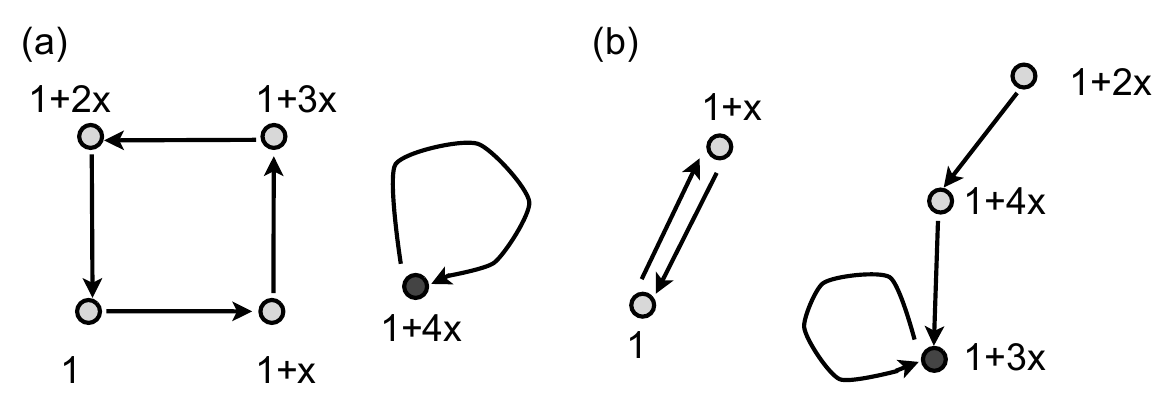}
\caption{Limit cycles in RG transformations for a classical fractal liquid with $f=1+x+x^{2}$ over $\mathbb{F}_{5}$. (a) $\mbox{RG}_{02}$. (b) $\mbox{RG}_{04}$.
} 
\label{fig_limit_cycle}
\end{figure}

Finally, consider $f=1+x+x^{2}$ over $\mathbb{F}_{5}$. We list some of its ground states as follows:
\ba{
&\begin{bmatrix}
1&\\
1&1&1\\
1&2&3&2&1\\
1&3&1&2&1& 3&1\\
1&4&0&1&4& 1&0&4&1
\end{bmatrix}
\begin{bmatrix}
1&1\\
1&2&2&1\\
1&3&0&0&3& 1\\
1&4&4&3&3 &4&4&1\\
1&0&4&1&0& 0&1&4&0&1
\end{bmatrix}\\
&\begin{bmatrix}
1&2\\
1&3&3&2\\
1&4&2&3&0&2\\
1&0&2&4&0&0&2&2\\
1&1&3&1&1&4&2&4&4&2
\end{bmatrix}
\begin{bmatrix}
1&3\\
1&4&4&3\\
1&0&4&1&2&3\\
1&1&0&0&2&1&0&3\\
1&2&2&1&2&3&3&4&3&3
\end{bmatrix}\\
&\begin{bmatrix}
1&4\\
1&0&0&4\\
1&1&1&4&4&4\\
1&2&3&2&4&2&3&4\\
1&3&1&2&4&3&4&4&2&4
\end{bmatrix}.
}
$\mbox{RG}_{02}$ generates the following limit cycle and fixed point:
\ba{
1 \rightarrow 1+x \rightarrow 1+3x \rightarrow 1+2x \rightarrow 1,\quad 1+4x \rightarrow 1+4x
}
where ground states are represented by $\gamma$'s. A transformation $\mbox{RG}_{04}$ leads to
\ba{
1 \rightarrow 1+x \rightarrow 1,\quad 1+2x \rightarrow 1+4x \rightarrow 1+3x \rightarrow 1+3x.
}
These sequences are shown in Fig.~\ref{fig_limit_cycle}. One may define renormalization function $\Gamma$ so that these fixed points and limit cycles are stable attractors.

\section{Quantum fractal liquid}\label{sec:QFL}


\subsection{$\mathbb{Z}_{2}$ spin liquid and polynomial}\label{sec:QFL1}

\begin{figure}[htb!]
\centering
\includegraphics[width=1.00\linewidth]{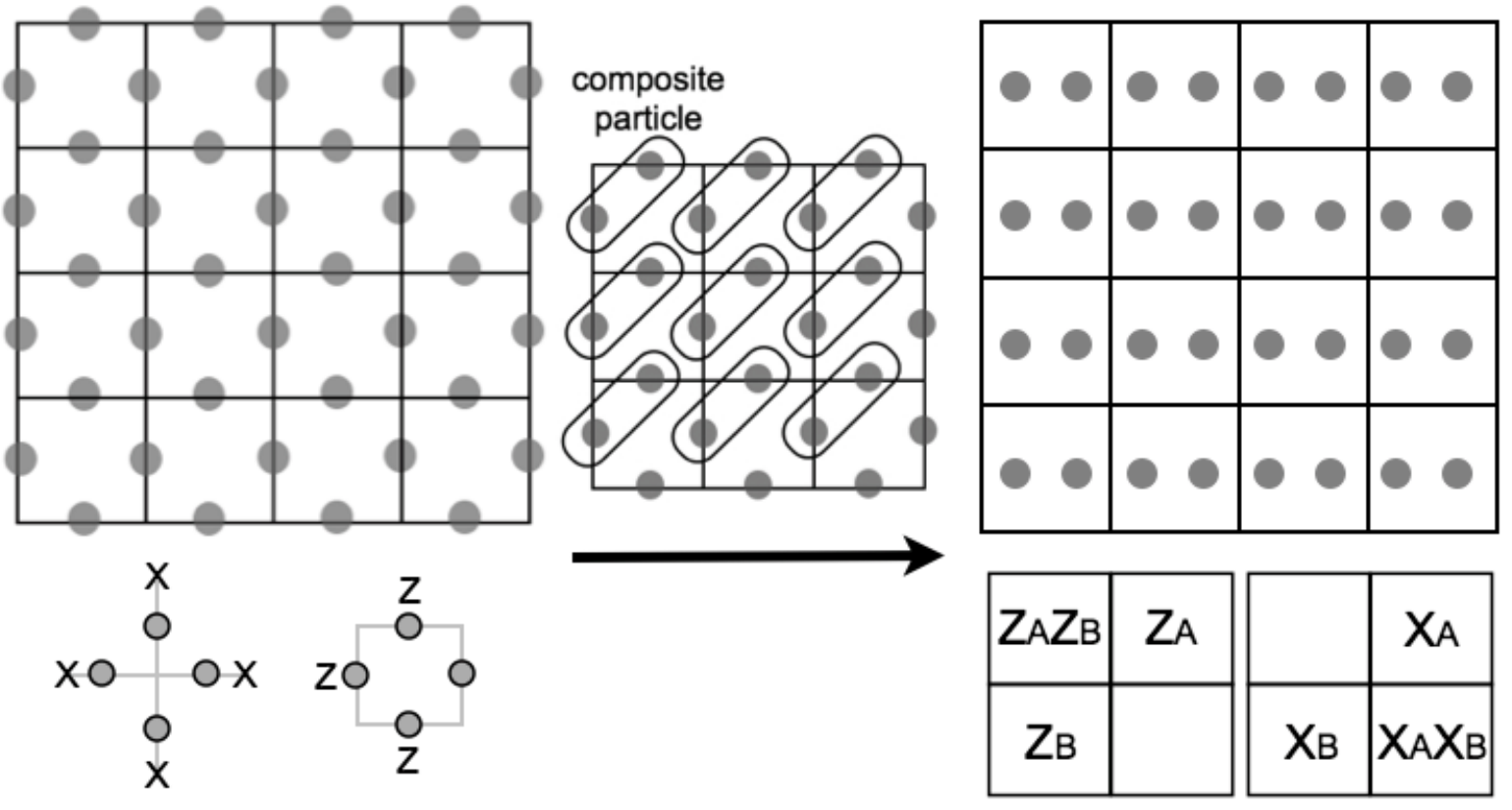}
\caption{Reduction of $\mathbb{Z}_{2}$ spin liquid. Two qubits are grouped into a composite particle which live inside each square. 
} 
\label{fig_Toric_poly}
\end{figure}

In this section, we present a general framework to construct a family of quantum fractal liquids which are condensation of fractal objects. We begin by representing $\mathbb{Z}_{2}$ spin liquid (the Toric code) by polynomials. Following Ref.~\onlinecite{Beni10b}, we group two qubits into a single composite particle (Fig.~\ref{fig_Toric_poly}) such that composite particles live on vertices of a square lattice:
\ba{
S^{(Z)}_{i,j} = 
\begin{bmatrix}
Z_{A}Z_{B}     ,& Z_{A}   \\
Z_{B}          ,&  I      
\end{bmatrix}\quad
S^{(X)}_{i,j} = 
\begin{bmatrix}
I ,& X_{A} \\
X_{B},& X_{A}X_{B} 
\end{bmatrix}
}
where each qubit inside a composite particle is labelled by $A$ and $B$. In polynomial representation, the parent Hamiltonian is 
\ba{
H = - \sum_{ij}Z \left(
\begin{array}{c}
x^{i}y^{j}(1+x) \\
x^{i}y^{j}(1+y)
\end{array}
\right) - \sum_{ij}X \left(
\begin{array}{c}
x^{i}y^{j}(1+y^{-1}) \\
x^{i}y^{j}(1+x^{-1})
\end{array}
\right)
}
where the upper (lower) row represents Pauli operators acting on $A$ ($B$). Interaction terms are translations of
\ba{
Z \left(
\begin{array}{c}
1+x \\
1+y
\end{array}
\right),\quad X \left(
\begin{array}{c}
1+y^{-1} \\
1+x^{-1}
\end{array}
\right)
}
In this form, it is immediate to see that $\mathbb{Z}_{2}$ spin liquid consists of a pair of one-dimensional ferromagnets $Z(1+x)$ and $Z(1+y)$ at its one-dimensional limits. 
 
Logical operators are
\ba{
&\ell^{(Z)}_{0} = Z\left( 
\begin{array}{c}
0 \\
1 + x + x^{2} + \cdots
\end{array}
\right)\\ 
&\ell^{(Z)}_{1} = Z\left( 
\begin{array}{c}
1+y+y^{2}+\cdots \\
0
\end{array}
\right)\\
&\ell^{(X)}_{0} = X\left( 
\begin{array}{c}
0 \\
1+y+y^{2}+\cdots 
\end{array}
\right)\\
&\ell^{(X)}_{1} = X\left( 
\begin{array}{c}
1 + x + x^{2} + \cdots\\
0 
\end{array}
\right)
}
It is worth representing them graphically as follows:
\ba{
\ell^{(Z)}_{0} = \begin{bmatrix}
Z_{B} & Z_{B} & \cdots &  Z_{B}&  Z_{B}\\
I& I & \cdots & I & I \\
\vdots & \vdots & \ddots  & \vdots &   \vdots \\
I& I & \cdots & I & I \\
I& I & \cdots & I &  I 
\end{bmatrix}\
\ell^{(Z)}_{0} = \begin{bmatrix}
Z_{A}& I & I &  \cdots & I\\
Z_{A}& I & I & \cdots & I\\
\vdots & \vdots  & \vdots &  \ddots & I\\
Z_{A}& I & I & \cdots& I \\
Z_{A}& I & I &  \cdots& I
\end{bmatrix}
}
and
\ba{
\ell^{(X)}_{0} = \begin{bmatrix}
X_{B}& I & I &  \cdots & I\\
X_{B}& I & I & \cdots & I\\
\vdots & \vdots  & \vdots &  \ddots & I\\
X_{B}& I & I & \cdots& I \\
X_{B}& I & I &  \cdots& I
\end{bmatrix}\
\ell^{(X)}_{1} = \begin{bmatrix}
X_{A} & X_{A} & \cdots &  X_{A} &  X_{A}\\
I & I & \cdots & I & I \\
\vdots & \vdots & \ddots  & \vdots &   \vdots \\
I & I & \cdots & I &  I \\
I & I & \cdots & I & I
\end{bmatrix}
}
One can see that logical operators commute with interaction terms by computing commutation polynomials. 

One can generalize construction of $\mathbb{Z}_{2}$ spin liquid. For arbitrary polynomials $\alpha(x,y,\cdots)$ and $\beta(x,y,\cdots)$, consider
\begin{align}
Z \left(
\begin{array}{c}
\alpha \\
\beta
\end{array}
\right),\quad X \left(
\begin{array}{c}
\bar{\beta} \\
\bar{\alpha}
\end{array}
\right)\label{eq:canonical}
\end{align}
where $\bar{\alpha}$ and $\bar{\beta}$ are duals of $\alpha$ and $\beta$ obtained by taking $x \rightarrow x^{-1}$, $y\rightarrow y^{-1}$, $\cdots$. Note that interaction terms commute with each other as their commutation polynomial is $\alpha \beta + \beta\alpha=0$ over $\mathbb{F}_{2}$. A parent Hamiltonian is 
\ba{
H = - \sum_{ij\cdots} Z \left(\begin{array}{c} x^{i}y^{j}\cdots \alpha \\ x^{i}y^{j}\cdots \beta \end{array} \right) - \sum_{ij\cdots} X \left(\begin{array}{c} x^{i}y^{j}\cdots \bar{\beta} \\ x^{i}y^{j}\cdots \bar{\alpha} \end{array} \right).
}
As for generalization to $\mathbb{F}_{p}$, we take $Z(\alpha,\beta)^{T}$ and $X(-\bar{\beta},\bar{\alpha})$ so that the commutation polynomial is $\alpha (- \beta) + \beta \alpha =0$. 

\subsection{Quantum fractal liquid}\label{sec:QFL2}

Consider a three-dimensional $L\times L \times L$ square lattice where two qubits live on each site with $L=2^{m}$ and periodic boundary conditions. Quantum fractal liquids have
\begin{align}
\alpha = 1 - f(x)y,\qquad \beta =1  - g(x)z
\end{align}
in Eq.~(\ref{eq:canonical}) where $f(x)$ and $g(x)$ are reversible polynomials over $\mathbb{F}_{2}$. More explicitly, interaction terms are translations of
\begin{align}
Z\left( 
\begin{array}{cc}
1 -f(x)y\\
1  - g(x)z
\end{array}
\right)
,\quad 
X
\left( 
\begin{array}{cc}
1 - \bar{g}(x) \bar{z} \\
1 - \bar{f}(x)\bar{y}
\end{array}
\right).
\end{align}
Interaction terms are characterized by a pair of fractal models $Z(1 -f(x)y)$ and $Z(1  - g(x)z)$. In this sense, quantum fractal liquids can be viewed as a coherent combination of a pair of classical fractal liquids living on $(\hat{x},\hat{y})$-plane and $(\hat{x},\hat{z})$-plane respectively. 

Logical operators of quantum fractal liquids have fractal shapes which are generated by polynomials $f(x)$ and $g(x)$:
\ba{
\tb{f}(x,y) &= 1 + fy + f^{2}y^{2} + \cdots \\ 
\bar{\tb{f}}(x,y) &= 1 + \bar{f}\bar{y} + \bar{f}^{2}\bar{y}^{2} + \cdots\\
\tb{g}(x,z) &= 1 + gz + g^{2}z^{2} + \cdots \\
\bar{\tb{g}}(x,y) &= 1 + \bar{g}\bar{z} + \bar{g}^{2}\bar{z}^{2} + \cdots.
}
Note $\tb{f}(x,y)$ lives on a $(\hat{x},\hat{y})$-plane while $\tb{g}(x,z)$ lives on a $(\hat{x},\hat{z})$-plane. Quantum fractal liquids have $k=2L$, and there are $2L$ of $Z$-type logical operators and $2L$ of $X$-type logical operators:
\ba{
&\ell^{(Z)}_{i}=
Z\left( 
\begin{array}{cc}
0 \\
x^{i}\tb{f}(x,y)
\end{array}
\right)\quad \ r^{(Z)}_{i}=
Z\left( 
\begin{array}{cc}
x^{i}\tb{g}(x,z)  \\
0 
\end{array}
\right)\\
&\ell^{(X)}_{i}=
X\left( 
\begin{array}{cc}
x^{i}\bar{\tb{f}}(x,y) \\
0  
\end{array}
\right)\quad 
r^{(X)}_{i}=
X\left( 
\begin{array}{cc}
0\\
x^{i}\bar{\tb{g}}(x,z)    
\end{array}
\right)
}
where $i=0,\cdots,L-1$. Therefore, $Z$-type logical operators have geometric shapes of $\tb{f}(x,y)$ and $\tb{g}(x,y)$ while $X$-type logical operators have geometric shapes of $\bar{\tb{f}}(x,y)$ and $\bar{\tb{g}}(x,y)$. 

To show that above operators are logical operators, we need to verify the following two things; a) they commute with interaction terms, b) they can be grouped into pairs of anti-commuting logical operators. One may see that logical operators commute with all the interaction terms by computing commutation polynomials. For instance, a commutation polynomial between $\ell_{0}^{(Z)}=Z(0,\tb{f}(x,y))^{T}$ and stabilizers $X(1-\bar{g}\bar{y},1-\bar{f}\bar{x})^{T}$ is given by $(1- fy)\tb{f}(x,y)=0$ for reversible $f$. Logical operators obey the following commutation relations:
\ba{\left\{
\begin{array}{cccccc}
\ell^{(Z)}_{0} ,& \cdots ,& \ell^{(Z)}_{L-1} ,& r^{(Z)}_{0} ,& \cdots ,& r^{(Z)}_{L-1} \\
r^{(X)}_{0} ,& \cdots ,& r^{(X)}_{L-1} ,& \ell^{(X)}_{0} ,& \cdots ,& \ell^{(X)}_{L-1}
\end{array}
\right\}.
}
Generalization to $\mathbb{F}_{p}$ is also possible.

To see that quantum fractal liquids are topologically ordered, we begin by showing that they are good quantum error-correcting code with $d \rightarrow \infty$ for $L\rightarrow \infty$ where $d$ is the quantum code distance of the ground state space. A standard way to prove this considers a bi-partition of the system into two complementary subsets $A$ and $B$ and uses the following bi-partition formula which holds for arbitrary stabilizer codes~\cite{Beni10}:
\begin{align}
g_{A}+g_{B}=2k
\end{align}
where $g_{A}$ and $g_{B}$ represent the number of independent logical operator supported inside $A$ and $B$ respectively. Let us assume $A$ to be a connected region with finite support. Then, its complementary subset $B$ accommodates some $(\hat{x},\hat{y})$-plane and $(\hat{x},\hat{z})$-plane where all the $2k$ independent logical operators can be supported. So, one has $g_{B}=2k$. This leads to $g_{A}=0$. Therefore, weights of logical operators are not finite (i.e. unbounded), and $d \rightarrow \infty$ for $L\rightarrow \infty$. 

For stabilizer Hamiltonians, being a quantum code ($d \rightarrow \infty$ for $L\rightarrow \infty$) automatically implies the presence of topological order with stability against local perturbations. Bravyi, Hastings and Michalakis~\cite{Bravyi10b} proved that frustration-free Hamiltonians with an ability of quantum error-correcting code have stability against local perturbations when Hamiltonians satisfy a certain condition, called TQO-2. Roughly speaking, TQO-2 states that locally computed density matrices are consistent with ground states which are computed globally. One can check that quantum fractal liquids satisfy TQO-2 by explicit calculations, and thus have stability against local perturbations. Recall that quantum fractal liquids have $2^{2L}$ ground states. Under a sufficiently small but finite local perturbations, the energy splitting among these ground states is always exponentially suppressed, and the energy gap between the ground states and excited states remains finite. 

We then discuss the number of degenerate ground states and its dependence on the system size. A key feature of quantum fractal liquids is that the number of logical qubits $k$ has a fairly sensitive dependence on the system size $\vec{L}=(L_{1},L_{2},L_{3})$. It turns out that the number of logical qubits $k$ is given by counting the dimension of solutions $\gamma$ satisfying the following equation:
\begin{align}
f(x)^{L_{2}} \gamma(x) = g(x)^{L_{3}}\gamma(x)= \gamma(x),\quad x^{L_{1}}=1\label{eq:number}
\end{align}
with $k = 2\dim \gamma$. For instance, with $f=x^{i}$ and $g=x^{j}$, one has $k=2\gcd (L_{1},iL_{2},jL_{3})$ where $k$ depends crucially on the system size $\vec{L}=(L_{1},L_{2},L_{3})$. In general, it is a very challenging task to write down an explicit form of $k(L_{1},L_{2},L_{3})$ for a given pair of $f(x)$ and $g(x)$. Yet, $k(L_{1},L_{2},L_{3})$ has a nice symmetry property under scale transformations:
\begin{align}
k(pL_{1},pL_{2},pL_{3})= p k(L_{1},L_{2},L_{3}).
\end{align}
This can be proven from discrete scale symmetries of polynomials over $\mathbb{F}_{p}$.

Ground states of quantum fractal liquids correspond to limit cycles under real-space RG transformations on an infinite lattice. To obtain RG transformations with meaningful attractors that do not flow to disordered states or trivial product states, one needs to apply some appropriate projection operators on sites that are to be coarse-grained. Below we present an example of such projections. Consider a pair of qubits at site $(i,j,\ell)$, denoted as $|\phi\rangle_{ij\ell} = |\phi_{A}\rangle_{ij\ell} \otimes |\phi_{B}\rangle_{ij\ell}$, and apply the following projections to a ground state:
\begin{align}
(I + Z_{A}^{\ell} \otimes Z_{B}^{j} )(I + X_{A}^{j} \otimes X_{B}^{\ell} )|\phi_{A}\rangle_{ij\ell} \otimes |\phi_{B}\rangle_{ij\ell}.
\end{align}
Note that projection operators commute with each other, and projections are applied only to sites $(i,j,\ell)$ with $j\not=0$ or $\ell\not=0$ modulo $2$. As a result, pairs of qubits on sites $(i,j,\ell)$ with $j=\ell=0$ modulo $2$ are completely decoupled from the rest. With some calculations, one notices that stabilizer generators for remaining sites $(i,j,\ell)$ with $j=\ell=0$ modulo $2$ are given by
\ba{
Z \left( 
\begin{array}{c}
\alpha^{2}\\
\beta^{2}
\end{array}
\right) \ Z \left( 
\begin{array}{c}
x\alpha^{2}\\
x\beta^{2}
\end{array}
\right) \ X \left( 
\begin{array}{c}
\bar{\beta}^{2}\\
\bar{\alpha}^{2}
\end{array}
\right) \ X \left( 
\begin{array}{c}
x\bar{\beta}^{2}\\
x\bar{\alpha}^{2}
\end{array}
\right)
}
and their translations that are generated by applications of $x^{2i'}y^{2j'}z^{2\ell'}$. (See Ref.~\onlinecite{Nielsen_Chuang} for transformations of stabilizer generators under projections). This corresponds to two copies of original quantum fractal liquids. Let us pick up sites with $(i,j,\ell)=(0,0,0)$ modulo $2$ and throw away sites with $(i,j,\ell)=(1,0,0)$ modulo $2$ via some arbitrary projections. Rescaling by $x^{2}\rightarrow x$, $y^{2}\rightarrow y$ and $z^{2}\rightarrow z$, stabilizer generators are $Z(\alpha,\beta)^{T}$ and $X(\bar{\beta},\bar{\alpha})^{T}$. So, this RG transformation maps a ground state of a quantum fractal liquid into some ground state which may be different from the original. One can keep track of how ground states flow under RG transformations by looking at polynomial representation of fractal logical operators. This can be analyzed in exactly the same way as ground states of classical fractal liquids. 

\section{Quasi-particles}\label{sec:no_string}

In this section, we discuss quasi-particle excitations and derive a necessary and sufficient condition for quantum fractal liquids to be free from string-like logical operators. Several examples of quantum fractal liquids are also studied, and the Cubic code is shown to be unitarily equivalent to a model of second-order quantum fractal liquid. 

\subsection{Criteria for no string}

We discuss properties of quasi-particle excitations in quantum fractal liquids. Without loss of generality, one can concentrate on excitations caused by Pauli-$Z$ type operators (phase errors) which flip $X$-type interaction terms. Following a treatment of classical fractal liquids, we represent positions of excitations as an excitation polynomial:
\ba{
E(x,y,z)=\sum_{i,j,\ell} c_{ij\ell}x^{i}y^{j}z^{\ell}\quad \mbox{over}\ \mathbb{F}_{p}
}
where $c_{ij\ell}=1$ means an excitation is present at $(i,j,\ell)$. We consider excitations caused by a Pauli operator $Z(e_{A},e_{B})^{T}$. They are given by a commutation polynomial between $(e_{A},e_{B})^{T}$ and $(-\bar{\beta},\bar{\alpha})^{T}$:
\ba{
E(x,y,z)= - e_{A}(1-gz) + e_{B}(1 -fy).
}

\begin{figure}[htb!]
\centering
\includegraphics[width=1.00\linewidth]{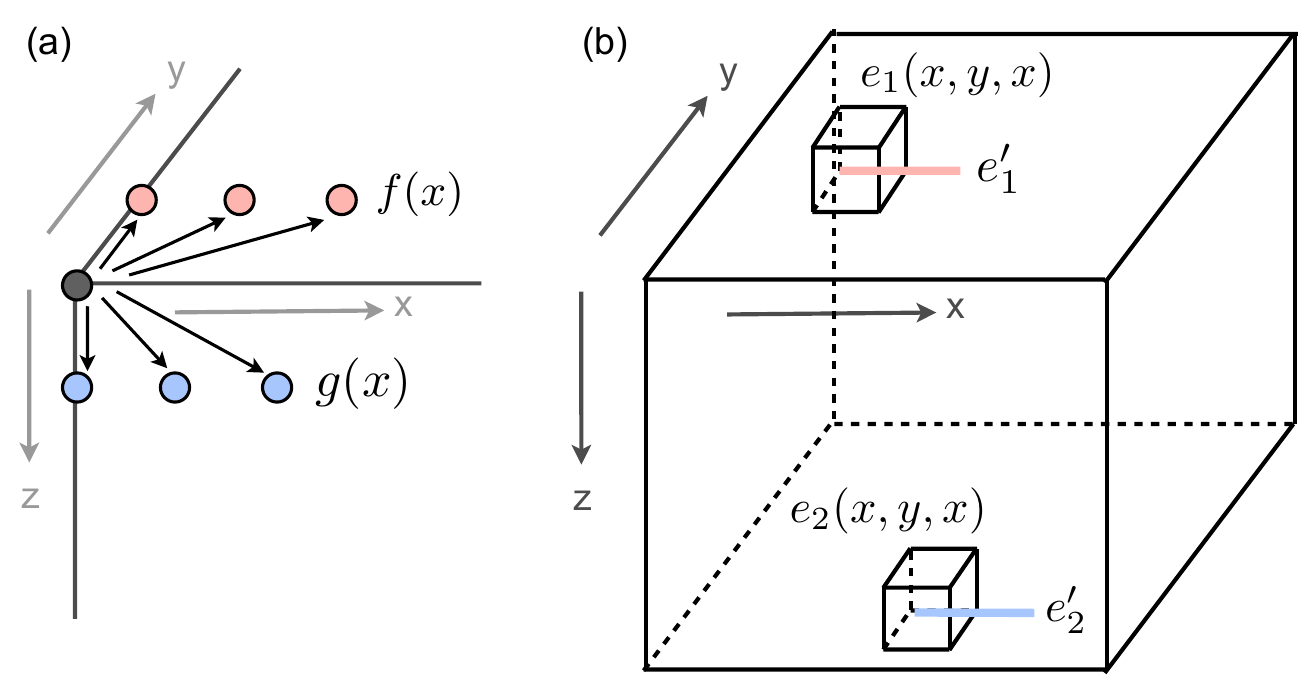}
\caption{(Color online) (a) Propagation of quasi-particles by $f(x)$ and $g(x)$ (b) A pair of localized excitations $e_{1}$ and $e_{2}$ with elongated excitations $e_{1}^{*}$ and $e_{2}^{*}$
} 
\label{fig_propagate}
\end{figure}

Excitations caused by a Pauli operator $Z_{0,0,0}^{(A)}$ ($e_{A}=1, e_{B}=0$) are given by
\begin{align}
E(x,y,z)=-(1-gz)
\end{align}
while excitations caused by a Pauli operator $Z_{0,0,0}^{(B)}$ ($e_{A}=0, e_{B}=1$) are given by
\begin{align}
E(x,y,z)=(1-fy).
\end{align}
So, if an isolated excitation is present at $(i,j,\ell)=(0,0,0)$, it will propagate to multiple excitations represented by $gz$ via an application of $Z_{0,0,0}^{(A)}$ (Fig.~\ref{fig_propagate}(a)). Similarly, it will propagate to multiple excitations represented by $fy$ via an application of $[Z_{0,0,0}^{(B)}]^{-1}$ (Fig.~\ref{fig_propagate}(a)). In general, for a single isolated excitation, $f(x)$ is applied when propagating in the $\hat{y}$ direction, and $g(x)$ is applied when propagating in the $\hat{z}$ direction. 

An analogy to cellular automaton becomes transparent by considering propagation of a one-dimensional excitation pattern $e(x)$, located at $j=\ell=0$. It will propagate in the $\hat{y}$ and $\hat{z}$ directions as follows:
\begin{align}
E(x,y,z)=e(x)f(x)^{j'}g(x)^{\ell'}y^{j'}z^{\ell'}.
\end{align}
This may viewed as time evolution of an initial condition $e(x)$, updated $j'$ times by $f(x)$ and $\ell'$ times by $g(x)$ respectively.

We first assume $f(x)=g(x)$, and consider propagation of $e(x)$: 
\begin{align}
E(x,y,z)=f(x)^{j' + \ell'} y^{j'}z^{\ell'}.
\end{align}
For $j'+\ell'=0$, excitations are single quasi-particles, and excitation energy remains finite. Therefore, quasi-particle excitations can propagate freely (without costing much energy) in the $\hat{y} - \hat{z}$ direction. This implies the presence of a string-like logical operator. Indeed, the following string-like operators are logical operators:
\ba{
Z\left( 
\begin{array}{cc}
(y+z)^{L-1} \\
(y+z)^{L-1} 
\end{array}
\right),\quad X\left( 
\begin{array}{cc}
(\bar{y}+\bar{z})^{L-1}\\
(\bar{y}+\bar{z})^{L-1}
\end{array}
\right).
}
Note that $(y+z)^{L-1}$ is a string-like polynomial extending in the $\hat{y}-\hat{z}$ direction. 

When do we have string-like logical operators? Without loss of generality, one can assume that $f(x)$ and $g(x)$ start from the origin, meaning that $f(x)$ and $g(x)$ have non-zero constant terms and have only positive powers. (Otherwise, we shift lattice positions). We say that $f(x)$ and $g(x)$ are \emph{algebraically related} when there exists some finite integers $c_{f}$ and $c_{g}$ such that
\begin{align}
f(x)^{c_{f}}=\mbox{const}\cdot g(x)^{c_{g}} \label{eq:related}
\end{align}
without considering periodic boundary conditions. Then, one notices that quasi-particles can propagate in the $c_{a}\hat{y}- c_{b}\hat{z}$ direction, and there exist string-like logical operators. 

It turns out that Eq.~(\ref{eq:related}) is a necessary and sufficient condition for the presence of string-like logical operators:
\ba{
&\mbox{No string-like logical operator} \quad \mbox{$\Leftrightarrow$} \\  &\mbox{$f(x)$ and $g(x)$ are not algebraically related}.
}
Our argument follows Ref.~\onlinecite{Bravyi11c}. Let us assume that a string-like logical operator exists. By taking a subpart of string-like logical operator, we may consider a pair of localized excitations $e_{1}(x,y,x)$ and $e_{2}(x,y,x)$ that are created at endpoints of string-like logical operators. We assume that excitations are contained in two cubic boxes of $w\times w \times w$ sites which are separated by $L_{\delta}$ with $L\gg L_{\delta}\gg w$. 

One can make quasi-particle excitations in $e_{1}(x,y,z)$ and $e_{2}(x,y,z)$ propagate by applying $f(x)$ and $g(x)$ such that they become elongated one-dimensional excitations whose lengths are $O(w)$ (see Fig.~\ref{fig_propagate}(b)). We denote polynomials corresponding to these elongated one-dimensional excitations as $e_{1}'$ and $e_{2}'$, and assume that $e_{1}'$ is at $j=\ell=0$ and $e_{2}'$ is at $j=j'$ and $\ell=\ell'$ where $|j'|+|\ell'|\sim O(L_{\delta})$. One may write $e_{1}'=e_{1}^{*}(x)$ and $e_{2}'=e_{2}^{*}(x)y^{j'}z^{\ell'}$. Then, one must have $e_{2}^{*} = -e_{1}^{*}f(x)^{j'}g(x)^{\ell'}$ as one can eliminate them by making $e_{1}'$ propagate and collide with $e_{2}'$. 

Since a pair of excitations is created by a string-like object (a subpart of string-like logical operator) with finite width, $f^{j'}g^{\ell'}$ must remain finite for large $j'$ and $\ell'$. This requires $f$ and $g$ to be algebraically related; otherwise, the size of $f^{j'}g^{\ell'}$ grows at least linearly as $|j'|$ and $|\ell'|$ grow. Therefore, the presence of string-like logical operators implies Eq.~(\ref{eq:related}).

\subsection{Several examples}\label{sec:QFL4}

Here, we study several examples of quantum fractal liquids. 

(a) We begin with a trivial case with $f=g=1$:
\ba{
Z\left( 
\begin{array}{cc}
1 + y\\
1  + z
\end{array}
\right)\qquad \mbox{over \ \ $\mathbb{F}_{2}$}
}
This is a stack of slices of two-dimensional $\mathbb{Z}_{2}$ model where each copy lives on a $(\hat{y},\hat{z})$-plane. It has pairs of string-like logical operators since $f=1$ and $g=1$ are generators of strings. Similarly, for $f=1$ and $g=1$ over $\mathbb{F}_{p}$ ($p>2$), the model is a stack of $\mathbb{Z}_{p}$ spin liquids with $p$-dimensional spins.

(b) For $f=x$ and $g=1$, stabilizer generators are given by 
\ba{
Z\left( 
\begin{array}{cc}
1 + xy\\
1  + z
\end{array}
\right)\qquad \mbox{over \ \ $\mathbb{F}_{2}$}.
}
This is a stack of slices of two-dimensional $\mathbb{Z}_{2}$ model, but each copy lives on a $(\hat{x} + \hat{y},\hat{z})$-plane. It has string-like logical operators, running in the $\hat{x}+\hat{y}$ direction. One can reduce this model to the model in (a) by a \emph{modular transformation} $x\rightarrow x$, $xy \rightarrow y$ and $z \rightarrow z$ which corresponds to a lattice distortion. 

(c) Let us look at an example with pairs of fractal logical operators and string-like logical operators:
\ba{
Z\left( 
\begin{array}{cc}
1 + (1+x+x^{2})y\\
1  + z
\end{array}
\right)\qquad \mbox{over \ \ $\mathbb{F}_{2}$}.
}
Geometric shapes of fractal logical operators are generated by $f=1+x+x^{2}$ (the Fibonacci model). Fractal logical operators live on a two-dimensional plane and string-like logical operators penetrate the two-dimensional plane.

(d) Some models do not have any logical qubits under periodic boundary conditions:
\ba{
Z\left( 
\begin{array}{cc}
1 + (1+x)y\\
1  + z
\end{array}
\right)\qquad \mbox{over \ \ $\mathbb{F}_{2}$}
}
since $f=1+x$ is not reversible over $\mathbb{F}_{2}$. When the model is defined with open boundary conditions, it has pairs of Sierpinski-like logical operators and string-like logical operators. 

(e) Consider
\ba{
Z\left( 
\begin{array}{cc}
1 + (1+x)y\\
1  +(1+x) z
\end{array}
\right),\qquad \mbox{over \ \ $\mathbb{F}_{2}$}.
}
The model has pairs of fractal logical operators, but has hidden string-like logical operators running in the $\hat{y} - \hat{z}$ direction. In fact, this model is unitarily equivalent to the following model
\ba{
Z\left( 
\begin{array}{cc}
1 + (1+x)y\\
1  +y^{-1} z
\end{array}
\right),\qquad \mbox{over \ \ $\mathbb{F}_{2}$}.
}

(f) Let us consider the cases without any string-like logical operators for $\mathbb{F}_{2}$:
\ba{
Z\left( 
\begin{array}{cc}
1 + (1+x+x^{2})y\\
1  +(1+x+x^{3}) z
\end{array}
\right)\qquad \mbox{over \ \ $\mathbb{F}_{2}$}
}
The model is free from string-like logical operators as $f=1+x+x^{2}$ and $g=1+x+x^{3}$ are algebraically unrelated over $\mathbb{F}_{2}$. Interaction terms are seven-body. 

(g) Next, let us consider a model over $\mathbb{F}_{p}$ ($p>2$):
\ba{
Z\left( 
\begin{array}{cc}
1 + (1+x)y\\
1  +(1+x^{2}) z
\end{array}
\right)\qquad \mbox{over \ \ $\mathbb{F}_{3}$}
}
The model is also free from string-like logical operators as $1+x$ and $1+x^{2}$ are algebraically unrelated over $\mathbb{F}_{3}$. Interaction terms are five-body. 

(h) The Cubic code can be viewed as a second-order quantum fractal liquid. In polynomial representation, one has
\ba{
Z\left(
\begin{array}{c}
1 + x + y + z\\ 
1 + xy + yz + zx
\end{array}
\right)\qquad \mbox{over \ \ $\mathbb{F}_{2}$}.
}
The model can be mapped to the following second-order quantum fractal liquid through local unitary transformations and modular transformations:
\ba{
Z \left( 
\begin{array}{c} 
1 + f(x)y \\ 
1+g_{1}(x)z + g_{2}(x)z^{2} 
\end{array}
\right)\qquad \mbox{over \ \ $\mathbb{F}_{2}$}
}
where 
\ba{
f(x)=1+x+x^{2},\quad g_{1}(x)=1+x,\quad g_{2}(x)=1+x+x^{2}.
}

Two-qubit gates can be characterized by a two-qubit Pauli operator $V=V_{1}\otimes V_{2}$. Consider an arbitrary two-qubit Pauli operator $U=U_{1}\otimes U_{2}$. A two-qubit gate generated by $V$ transforms $U$ as follows:
\begin{align}
U_{1}\otimes U_{2} \ \rightarrow \ U_{1}V_{1}^{c_{2}} \otimes U_{2}V_{2}^{c_{1}} \label{eq:2qubit}
\end{align}
where $U_{1}V_{1}= (-1)^{c_{1}}V_{1}U_{1}$ and $U_{2}V_{2}= (-1)^{c_{2}}V_{2}U_{2}$. For instance, with $V=X_{1}\otimes X_{2}$, one has the following transformations:
\ba{
\left\{
\begin{array}{cc}
Z_{1} ,& Z_{2} \\
X_{1} ,& X_{2}
\end{array}
\right\} \ \rightarrow \ 
\left\{
\begin{array}{cc}
Z_{1}X_{2} ,& X_{1}Z_{2} \\
X_{1} ,     & X_{2}
\end{array}
\right\}.
}
These two-qubit gates may be viewed as generalizations of control-$Z$ operation. One may see that transformations in Eq.~(\ref{eq:2qubit}) preserve commutation relations by direct calculations, and thus are indeed unitary transformations.

Let us apply these two-qubit gates to a canonical model with $Z(\alpha,\beta)^{T}$ and $X(\bar{\beta},\bar{\alpha})^{T}$. We think of applying a sequence of two-qubit gates, characterized by $X_{A}\otimes Z_{B}$, on neighboring sites in the $\hat{x}$ direction. Then, one has the following transformations:
\ba{
Z\left(
\begin{array}{c}
\alpha \\
\beta
\end{array}
\right) \rightarrow 
Z\left(
\begin{array}{c}
\alpha \\
\beta + \alpha x
\end{array}
\right)\quad 
X\left(
\begin{array}{c}
\bar{\beta} \\
\bar{\alpha}
\end{array}
\right)  \rightarrow 
X\left(
\begin{array}{c}
\bar{\beta} + \bar{\alpha} \bar{x} \\
\bar{\alpha}
\end{array}
\right)
}
which correspond to $\alpha \rightarrow \alpha$ and $\beta \rightarrow \beta + x\alpha$. Note that these two-qubit gates can be applied simultaneously as they commute with each other. By generalizing this transformation, the following transformations are allowed:
\begin{align}
\alpha \rightarrow \alpha,\qquad \beta \rightarrow \beta + x^{i}y^{j}z^{\ell}\alpha
\end{align}
where $i,j,\ell$ are some finite integers. 

For the Cubic code ($\alpha = 1 + x + y + z$, $\beta = 1 + xy + yz + zx$), we apply two-qubit gates $(\alpha ,\beta) \rightarrow (\alpha,\beta + x\alpha)$, modular transformations $(x,y,z)\rightarrow (x, yz^{-1},z)$, shifting of lattice sites in the $\hat{z}$ direction and two-qubit gates $(\alpha,\beta)\rightarrow (\alpha + \beta, \beta)$:
\ba{
&\left(
\begin{array}{c}
1 + x+ y+z \\
1 + xy + yz + zx
\end{array}
\right) \ \rightarrow \ 
\left(
\begin{array}{c}
1 + x+ y+z \\
1 + x + x^{2} + yz
\end{array}
\right) \ \rightarrow \\ 
&\left(
\begin{array}{c}
1 + x + yz^{-1} + z\\
1 + x + x^{2} + y
\end{array}
\right) 
\ \rightarrow \
\left(
\begin{array}{c}
y + (1+x)z +  z^{2}\\
1 + x + x^{2} + y
\end{array}
\right) 
\ \rightarrow \\
&\left(
\begin{array}{c}
(1+x+x^{2}) + (1+x)z +  z^{2}\\
1 + x + x^{2} + y
\end{array}
\right).
}

\section{Outlook}\label{sec:conclusion}

In this paper, we have presented general construction of classical and quantum fractal liquids and demonstrated that they have exotic physical properties beyond theory with continuous scale invariance. We hope that our construction and theory of fractal spin liquids will provide a stepping stone toward complete understanding and classification of quantum phases of matter. 

Quasi-particle excitations arising in quantum fractal liquids are all Abelian. Topological order arising in quantum fractal liquids is non-chiral. It is unclear to what extend our results may generalize to chiral topological phases, non-Abelian topological phases and symmetry protected topological phases. Effective field theoretical descriptions of classical and quantum fractal liquids are currently not known. It may be interesting to analyze how classical and quantum fractal liquids behave under RG transformations in the language of matrix and tensor product state representations.~\cite{Vidal07, Verstraete04b, Gu08, Evenbly09} An underlying difficulty in physically realizing the Sierpinski triangle model lies in the fact that the model has three-body interaction terms. Yet, one may simulate three-body interactions easily by using hopping particles.~\cite{Ocko11b} Construction of quantum fractal liquids builds on cellular automaton with linear update rules, and non-linear extension remains as an open problem. 

Another important motivation behind this paper concerns quantum information storage capacity of discrete spin systems. There is a well-known bound on the amount of quantum information that can be stored reliably in a given volume of discrete spin systems which are supported by gapped local Hamiltonians.~\cite{Bravyi10} However, all the previously known systems were far below this theoretical bound, and it remains open whether there exists a gapped spin system that saturates this bound. We have solved a classical version of this problem~\cite{Beni11b} by proving that a family of Sierpinski-type classical fractal liquids asymptotically saturate the classical information storage capacity bound. With this success in hand, we hope that quantum fractal liquids, which are natural generalization of classical fractal liquids, also asymptotically saturate the quantum information storage capacity bound. Analysis on coding properties of quantum fractal liquids is an important open problem which may lead to discovery of capacity saturating quantum error-correcting codes. 

\section*{Acknowledgment}

I thank Peter Shor, John Preskill, Xiao-Gang Wen and Shintaro Takayoshi for helpful discussion and comments. I acknowledge fruitful discussion with Oliver Buerschaper on RG transformations. I thank Jeongwan Haah, Spiros Michalakis and Alex Kubica for countless discussions at Caltech. I am supported by David and Ellen Lee Postdoctoral fellowship. I acknowledge funding provided by the Institute for Quantum Information and Matter, an NSF Physics Frontiers Center with support of the Gordon and Betty Moore Foundation (Grants No. PHY-0803371 and PHY-1125565).

\end{document}